# Pulsed laser deposition of $Sb_2S_3$ films for phase-change tunable nanophotonics

PETER KEPIČ[1,2], PETR LIŠKA[1,2], BEÁTA IDESOVÁ[1,2], ONDŘEJ CAHA[3], FILIP LIGMAJER[1,2,*], AND TOMÁŠ ŠIKOLA[1,2]

[1]*Central European Institute of Technology, Brno University of Technology, 612 00 Brno, Czech Republic*
[2]*Institute of Physical Engineering, Faculty of Mechanical Engineering, Brno University of Technology, 616 69 Brno, Czech Republic*
[3]*Department of Condensed Matter Physics, Masaryk University, Brno, Czech Republic*
*\*filip.ligmajer@vutbr.cz*

**Abstract:** Non-volatile phase-change materials with large optical contrast are essential for future tunable nanophotonics. Antimony trisulfide ($Sb_2S_3$) has recently gained popularity in this field due to its low absorption in the visible spectral region. Although several $Sb_2S_3$ deposition techniques have been reported in the literature, none of them was optimized with respect to the lowest possible absorption and largest optical contrast upon the phase change. Here, we present a comprehensive multi-parameter optimization of pulsed laser deposition of $Sb_2S_3$ towards this end. We correlate the specific deposition and annealing parameters with the resulting optical properties and propose the combination leading to films with extraordinary qualities ($\Delta n = 1.2$ at 633 nm). Finally, we identify crystal orientations and vibrational modes associated with the largest change in the refractive index and propose them as useful indicators of the $Sb_2S_3$ switching contrast.

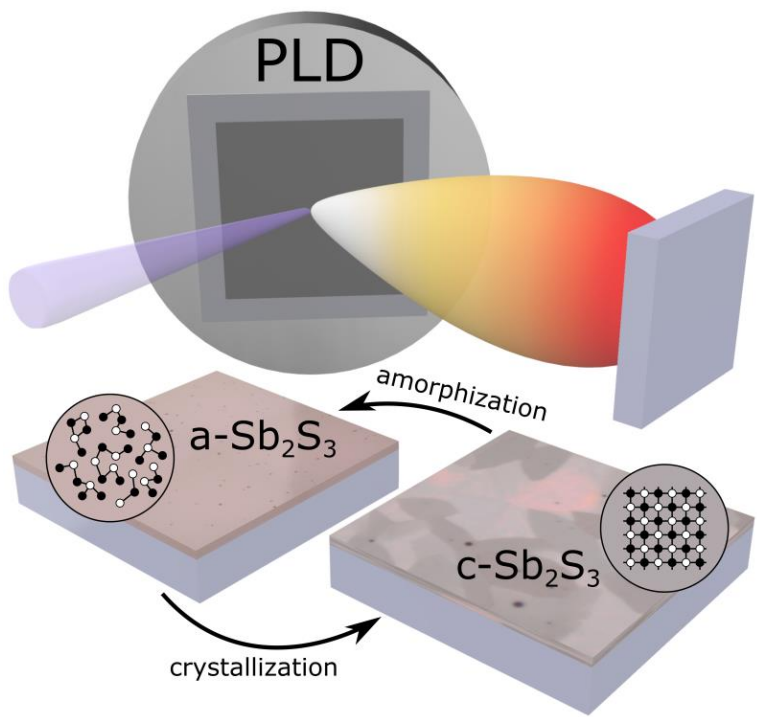

## 1. Introduction

Phase-change materials with solid-solid phase transition have been known for their use in data storage for more than three decades [1]. Nowadays, their applications shift towards tunable nanophotonics: tunable beam steering, tunable focusing, or tunable amplitude filters have been demonstrated using non-volatile or volatile phase-change materials [2-4]. The most popular non-volatile phase-change materials are germanium antimony telluride (GST) compounds. This popularity stems from their substantial refractive index contrast during the transition ($\Delta n \approx 3.5$ at 1550 nm), fast switching speed ($\leq$ 1 ns), and long endurance (billions of cycles) [5,6]. However, their use is limited to the near-infrared or infrared spectral region because of their strong absorption at shorter wavelengths.

Antimony trisulfide ($Sb_2S_3$) is a suitable replacement for GST compounds, as it has near-zero absorption in the visible region while still offering a relatively large refractive index contrast between its amorphous and crystalline phases ($\Delta n \approx 1.1$ at 633 nm) [7]. The crystallization at around 250 °C and amorphization at around 540 °C are attainable by external heating or laser irradiation [8]. Note that the low crystallization temperature renders the material more stable in the air than GST [7,9]. Unfortunately, these advantages come at the cost of a lower endurance ($\approx$ 1000 cycles) [10,11], which arises from the gradual degradation of local atomic structure at the nucleation sites [12]. Another drawback is the highly polycrystalline character of $Sb_2S_3$ after crystallization and its large inherent optical anisotropy [13–15], which can both negatively impact the desired functionalities of tunable nanostructures made of this material. While understanding the properties of individual crystal grains in $Sb_2S_3$ films and the interplay of their structural and optical properties are still not fully clear [8], tunable color pixels and integrated waveguide switches have already been demonstrated [7,10,16–21]. Although $Sb_2S_3$ films in these nanophotonic applications were deposited using sputtering techniques, other techniques, such as evaporation, chemical bath deposition, atomic layer deposition, and pulsed laser deposition, have been reported as well [22–26]. Concerning optical, structural, and compositional properties of $Sb_2S_3$ films, these techniques could offer a an adequate or even superior replacement for sputtering. However, their thorough optimization, especially focused on the contrast of optical properties upon the $Sb_2S_3$ transition, is lacking in the literature.

This article focuses on optimizing the pulsed laser deposition (PLD) of $Sb_2S_3$ thin films of various thicknesses. In the first part, we look at the influence of the deposition laser frequency, its fluence, and deposition gas pressure on the morphological, compositional, and optical properties of $Sb_2S_3$. In particular, we expose the close relationship between stoichiometry, optical absorption, and refractive index contrast during the transition. We show that the contrast upon the phase transition can be significantly improved by varying the pre-deposition pressure and the crystallization ramp rate. Lastly, we investigate whether the optical properties of amorphous and crystalline phases of $Sb_2S_3$ films depend on their thickness. Using the refractive index contrast as the main metric again, we present its relations with the thickness, stoichiometry, and crystal structure. Moreover, we even identify crystal orientations and Raman modes associated with the most considerable refractive index contrast and propose them as useful indirect indicators of the $Sb_2S_3$ switching contrast.

## 2. Methods

### 2.1 Thin film deposition

The $Sb_2S_3$ films were deposited on top of 10×10 mm$^2$ silicon substrates (Siegert Wafer, (100) orientation, P-doped) by the TSST PLD system using a 248 nm KrF laser. Before the first deposition, the stoichiometry of an $Sb_2S_3$ sputtering target (99.99% purity, Testbourne B.V.) was verified by energy-dispersive X-ray spectroscopy (EDS). A 3×15 mm$^2$ laser mask provided a 0.55×1.95 mm$^2$ ablation spot. The 16×16 mm$^2$ area of the target was pre-ablated before each deposition (5120 pulses at 10 Hz). Each substrate was cleaned before each deposition in an ultrasound bath of acetone (2 min), isopropyl alcohol (2 min) and deionized water (1 min) and baked on a 150 °C hotplate for 1 min to evaporate the residual water. The deposition was carried out from a 12×12 mm$^2$ central area of the target, while its distance from the sample was fixed at 50 mm. The pre-deposition pressure (before each deposition) was around 2×10$^{-3}$ mTorr if not stated otherwise. The remaining parameters, such as the deposition laser frequency, fluence, Ar deposition pressure, and the number of pulses, were optimized as discussed in the main text below.

Protective Al$_2$O$_3$ thin films, with a thickness of around 15 nm, were deposited using an atomic layer deposition system (Ultratech/CambridgeNanoTech Fiji 200) with a trimethylaluminium precursor (99.999% Al basis) and water as a co-reactant. The temperature of the chamber and sample was 100 °C. The carrier and purge gas were argon with flow rates of 30 sccm and 100 sccm, respectively. The film was deposited in 140 cycles.

## 2.2 Morphological characterization

The composition of Sb$_2$S$_3$ films was characterized by X-ray photoelectron spectroscopy (XPS) and EDS: For XPS, we used the Kratos Axis Supra model with a background pressure around 7×10$^{-5}$ mTorr and combined Al/Ag anode (1486.6 / 2984.3 eV). All samples were measured ex situ immediately after the deposition (before the protective film). The process of fitting the XPS spectra is described in Fig. S3. For EDS, we used the Bruker XFlash 5010 spectrometer integrated into a Tescan LYRA3 microscope, with a voltage of 15 kV, a current around 1.2 nA, and a 9 mm working distance. For each sample, we averaged 100 points on three different 2×2 mm$^2$ areas. The measured spectra were analyzed using the PB-ZAF method.

## 2.3 Compositional characterization

Before depositing the protective film, the morphology of each amorphous Sb$_2$S$_3$ film was investigated using an atomic force microscope (Bruker Dimension Icon). The scanned area was 5×5 μm$^2$, with a resolution of 512×512 pixels and a scanning speed of 1 Hz. The root mean square (RMS) roughness was obtained from AFM images using the Gwyddion software [27]. The surface of amorphous and crystalline Sb$_2$S$_3$ films with the protective Al$_2$O$_3$ film was also investigated using a scanning electron microscope (Tescan LYRA3).

## 2.4 Structural characterization

Structural characterization of all samples was performed using Raman spectroscopy and X-ray diffraction (XRD): For Raman spectroscopy, we used a confocal Witec Alpha 300R system with a 100× objective (NA 0.9, WD 0.31 mm). A 633 nm excitation laser was chosen to provide the low absorption of light in the films and thus prevent them from their local crystallization. Specifically, for the amorphous phase of Sb$_2$S$_3$ films, we used an 1800 g/mm grating with a 400 cm$^{-1}$ spectral center, 1 mW laser power, 1 s integration time, and 60 accumulations. For the crystalline phase of films below 264 nm thickness, we used the same grating, 6 mW laser power, 300 s integration time and 1 accumulation. For the films of 264, 386 and 493 nm thickness, the power was adjusted to 4, 3 and 2 mW, respectively, because their overall absorption was more significant due to their greater thickness. As each sample is naturally polycrystalline, we averaged the Raman spectra of 4 differently-looking crystal grains for each sample to get a response as close as possible to that one of polycrystalline materials, similar to ellipsometry and XRD measurements, which collect data averaged from larger areas.

XRD measurements were carried out using a Rigaku SmartLab 9 kW system with a Cu Kα (1.542 Å) source, operated at 45 kV and 150 mA. For all measurements, we fixed the incident beam at 0.5° and collected data from 10° to 80° by 0.1° step with a D/teX Ultra detector. The attenuator was set to automatic, and the measured data were normalized to the maximum peak. The measured peaks were identified by Rigaku PDXL software, connected to the ICDD PDF-2 database.

## 2.5 Optical characterization

Optical characterization of thin films was done with a spectroscopic ellipsometer (J. A. Woollam V-VASE). The measurement was carried out using an adjustable compensator, 1000 µm monochromator slit, 50°, 60° and 70° incident angles, and 0.20–4.00 eV spectral range with 0.04 eV step.

## 3. Deposition optimization

One of the goals in phase-change tunable nanophotonics is to find a material whose refractive index contrast between two phases is as large as possible while the absorption coefficient stays low (ideally identical in both phases). The most promising phase-change material for visible tunable nanophotonics in this respect is $Sb_2S_3$. With the 2.05 eV bandgap [20], the amorphous $Sb_2S_3$ films, on average [7,10,11,16–18], exhibit almost negligible (≤ 0.02) optical absorption at wavelengths above 633 nm. The average refractive index contrast between its amorphous and crystalline phase goes up to 0.8 while keeping the absorption contrast as low as 0.7 in the middle of the visible region [7,10,11,16–18]. Such optical properties are hardly achievable with the most popular GST compounds ($k_{amorphous} \geq 1.5$, $\Delta n \approx 0.5$ at 633 nm) [5].

While most of the $Sb_2S_3$ films applied to nanophotonics have been deposited by sputtering techniques, there are also other well-described techniques [22–26]. However, for these techniques, we are unaware of any reports elucidating the influence of the deposition parameters on the resulting optical contrast. One such deposition technique is PLD [25,26], which we have chosen to investigate, as it is also a fast technique (≈ 2.6 Å/s) in which all deposition parameters can be well controlled.

We first investigated how the most fundamental parameters of PLD (laser frequency, laser fluence, and Ar deposition pressure) influence morphology, stoichiometry, and optical properties of amorphous $Sb_2S_3$ films: The laser frequency had no observable impact (see Fig. S1), so we decided to fix it at the maximum possible value of 10 Hz. Then, we deposited $Sb_2S_3$ films using various combinations of 3 different laser fluences and 10 values of Ar pressure and characterized them accordingly (see Methods). The results are presented as 2D maps with the laser fluence and Ar pressure axes. Data between measured points are linearly interpolated. First, we investigated the film roughness, which is important for the fabrication of nanostructures, by ex-situ AFM (see Fig. 1(a)). The smoothest films with RMS values below 1 nm were grown at Ar pressures below 20 mTorr. In the region of 20 - 40 mTorr of Ar, we observed a significant roughness increase (up to RMS of 2.5 nm) associated with the formation of Sb crystals on the film surface (see Fig. S2(a-b)). The surfaces of $Sb_2S_3$ films deposited at pressures higher than 40 mTorr were generally smoother, reaching RMS values around 1.3 nm (see Fig. S2(c)). Regarding stoichiometry, to find atomic concentration ratios of S and Sb atoms, we investigated the ratio between S $2p^{3/2}$ and Sb $3d^{5/2}$ peaks [10] of the XPS spectra by a fitting procedure described in Fig. S3. The results shown in Fig. 1(b) prove that the desired $Sb_2S_3$ stoichiometry of 1.5:1 can be generally obtained at lower fluences and higher Ar pressures. Interestingly, we can associate the formation of Sb crystals described in Fig. 1(a) with the transition between the 1:1 and 1.5:1 stoichiometric ratios. The almost identical trend can be seen in the Sb, $Sb_2O_3$ and $Sb_2S_3$ components of the Sb $3d^{5/2}$ peak of the XPS spectra and also in an additional EDS measurement, which are both presented in Fig. S4. The first experiments indicate that the smoothest films with good stoichiometry are generally produced at low laser fluences and Ar pressures above 50 mTorr. Further, the influence of deposition parameters on the optical properties of the films will be discussed.

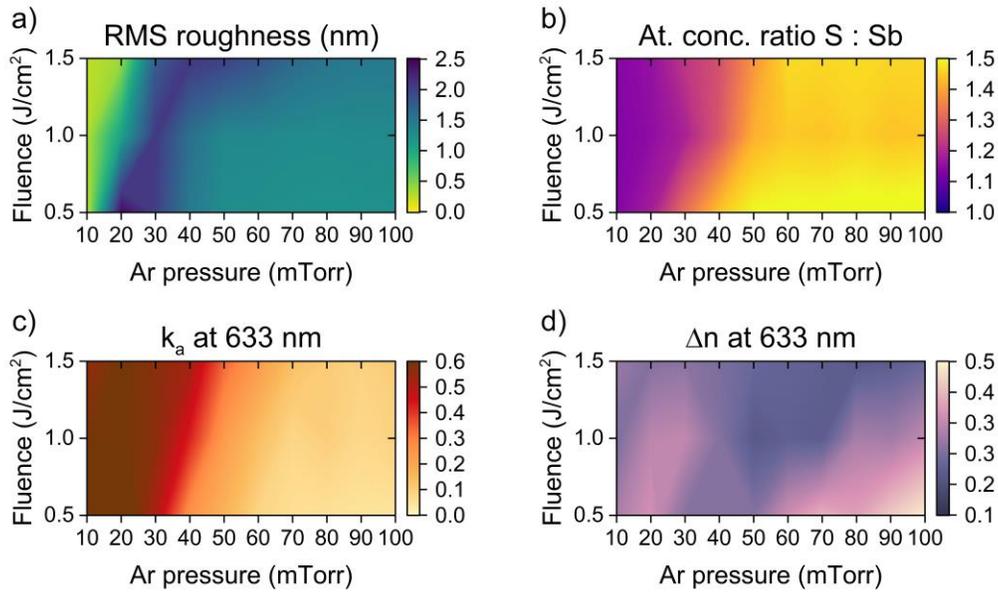

Fig. 1. a) RMS roughness determined from 5×5 μm² AFM images, (b) Concentration ratio of S and Sb atoms determined from S $2p^{3/2}$ and Sb $3d^{5/2}$ peaks of fitted XPS spectra, (c) absorption coefficient of amorphous $Sb_2S_3$ films at 633 nm wavelength, and (d) refractive index contrast between crystalline and amorphous $Sb_2S_3$ films at 633 nm wavelength, all plotted as 2D functions of the deposition laser fluence and Ar pressure. The data in (c) and (d) were obtained from modeled dielectric functions of amorphous and crystalline $Sb_2S_3$ films fitted to measured ellipsometric spectra.

As $Sb_2S_3$ films tend to lose sulfur during the crystallization process [28], we protected them by approx. 15 nm of $Al_2O_3$ (following Ref. [11], see Methods). Subsequently, we performed spectroscopic ellipsometry of amorphous $Sb_2S_3$ films (see Methods), fitted the ellipsometric data (Fig. S5), extracted the dielectric functions (see Fig. S7), and plotted the absorption coefficient at 633 nm as a 2D function of the deposition parameters (Fig. 1(c)). When compared with Fig. 1(b), it is obvious that the absorption is strongly correlated with stoichiometry. In detail, the stoichiometry closer to the ideal $Sb_2S_3$ corresponds to lower absorption of amorphous $Sb_2S_3$ films, reaching down to 0.03, again at low laser fluences and high Ar pressures. Note that a similar trend was also observed for other visible wavelengths (see Figs. S7 and S8(a)). After a complete analysis of all the amorphous films, their crystallization was done by placing them for 2 minutes directly onto a hotplate pre-heated to 300 °C. The average temperature ramp rate measured by a thermocouple was approx. 160 °C/min. Subsequently, the crystallized films were re-measured by spectroscopic ellipsometry, the spectra were fitted (Fig. S5), and from the obtained dielectric functions (Fig. S7), we extracted the refractive index at 633 nm. The refractive index contrast between the amorphous and crystalline films (Δn) was then plotted again as a 2D function of the deposition parameters (Fig. 1(d)). While the Δn gets larger at lower fluences and higher Ar pressures, there is an anomaly between 30-70 mTorr of Ar pressure. As the refractive index before the crystallization followed the absorption trend (see Fig. S8(b)), the discrepancy should be associated with the crystallization process (see Fig. S8(c)). However, as the discrepancy is not visible in the absorption (Fig. 1(c)), it is a more complex problem that we decided not to focus on. While all properties seem to be improving towards larger pressures, we observed that films deposited at pressures above 100 mTorr exhibit more than 15% inhomogeneity in thickness already over a 10 mm distance. This effect most probably arises from the influence of Ar on the volume density (shape) of deposition plasma [29] and limits large-scale deposition. Therefore, we decided to stop at the edge of the investigated area, look into the structural properties of the most promising film deposited at 0.5

J/cm$^2$ and 100 mTorr and try to investigate other deposition and crystallization parameters that might further improve the optical properties of this film.

When observing the amorphous film in an optical microscope (Fig. 2(a), orange), we saw that even for deposition parameters that lead to almost the perfect stoichiometry, we could not fully eliminate the unwanted Sb crystals (black dots) segregated on top of our films. Due to their size and density, they, fortunately, represent only minor limitations for the future fabrication of nanostructures. Looking at the real ($n$) and imaginary ($k$) parts of the refractive index in Fig. 2(b), we can claim that our amorphous film agrees well with the amorphous $Sb_2S_3$ films reported in the literature [7]. If we look at the XRD pattern in Fig. 2(c), we can see three broad peaks that correspond to the amorphous film and two additional peaks at 29° and 51.5° that can be associated with the (012) and (202) preferential orientation of Sb crystals, respectively. This supports our previous identification of Sb crystals on the surface. The amorphous character of our $Sb_2S_3$ film is also confirmed by two broad Raman peaks around 100 cm$^{-1}$ and 300 cm$^{-1}$ in Fig. 2(d), ascribed to Sb–S and S=S vibrational modes, respectively [10].

When the sample is heated for 2 minutes at 300 °C on a hot plate, we can observe changes in optical properties (Fig. 2(b), blue). These optical changes result from the crystallization process, as indicated by the XRD pattern showing orthorhombic $Sb_2S_3$ peaks (Fig. 2(c)). Note that the high density of these peaks (often presented as an indicator of high $Sb_2S_3$ quality [30]) also implies a highly polycrystalline character of our film. This could result in varying optical properties at the nanoscale and, therefore, should be considered when $Sb_2S_3$ nanostructures are fabricated. Similarly, from the Raman spectra in Fig. 2(d), we can distinguish 12 active Raman peaks of the crystalline $Sb_2S_3$ that strengthen the claim about the crystalline phase transition [30, 31]. All these results show that our PLD film has structural qualities comparable to the films reported in the literature [7,10]. However, we observed that the film integrity was disrupted during the crystallization (Fig. 2(a), blue), which is undesired for future applications. An abrupt increase in the temperature by putting samples directly onto the hotplate at 300 °C probably caused a rapid thermal expansion of the silicon substrate before the $Sb_2S_3$ film could be fully crystallized. This led to the formation of cracks and voids in the film. We decided to verify this hypothesis by slowing down the temperature ramp rate through a continuous increase in the temperature.

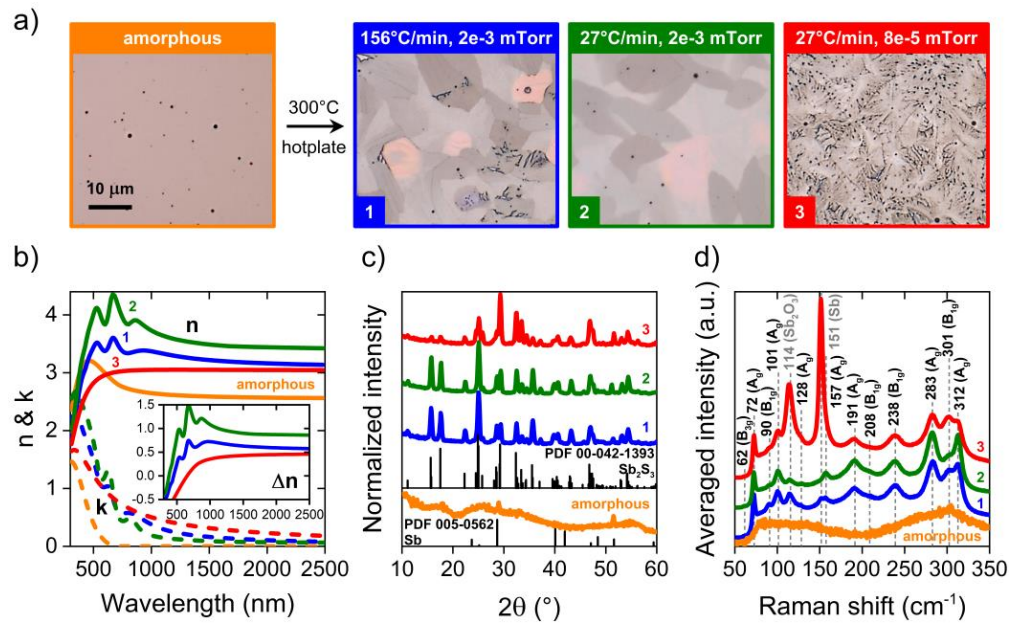

Fig. 2. a) Optical microscopy images of an amorphous $Sb_2S_3$ film (deposited at 10 Hz, 0.5 J/cm$^2$, 100 mTorr, 5000 pulses) and three crystalline $Sb_2S_3$ films crystallized at different conditions. The crystallization temperature ramp rates and pre-deposition pressures are listed in the images. b) Real (n) and imaginary (k) parts of the measured refractive index, (c) normalized XRD pattern, and (d) Raman spectra of the films shown in (a). The inset in (b) shows the refractive index contrast between the amorphous and relevant crystalline phases in (b). The line colors correspond to the image frame colors in (a).

Using another amorphous film from the same optimized batch, we tried slower crystallization with an average temperature ramp rate of 27 °C/min. As shown in Fig. 2(a) (green), the resulting crystallized film was indeed without cracks. This integrity improvement led to an almost two-fold increase in the refractive index contrast (inset in Fig. 2(b)). The absorption coefficient increased by 30% in the visible, while at the same time it decreased by 50% in the near-infrared region. Moreover, this slower-crystallized film exhibits a 1.5 times larger value of the refractive index contrast than the average of Refs. [7,10,16–21] (see Fig. S10). Structurally, the slower-crystallized film (green) had almost identical XRD (Fig. 2(c)) and Raman (Fig. 2(d)) peaks as the one with cracks (blue). As the crystallized films were structurally almost identical, but their real (n) and imaginary (k) parts of the refractive index changed, we believe this optical improvement came from the film integrity. While integrity heavily impacts the refractive index (through an effective medium with voids), it has minimal impact on the structural properties resulting from the atomic configuration. To avoid possible complications with integrity, we recommend gradual annealing when the crystallization of the $Sb_2S_3$ film is induced at higher temperatures.

For some deposition techniques, the quality of the vacuum before the deposition plays a key role in the quality of the deposited film. We, therefore, tested the influence of the pre-deposition pressure (i.e., before introducing Ar deposition gas) on the refractive index contrast during the phase transition of $Sb_2S_3$. We pre-ablated the PLD target, inserted the sample and waited for the pre-deposition pressure to drop down to $8\times10^{-5}$ mTorr instead of $2\times10^{-3}$ mTorr as before. The deposited amorphous film had identical optical and structural properties as the film deposited at the higher pre-deposition pressure. However, when crystallized, the film exhibited a highly porous character with many open pores reaching down to the substrate (see Fig. 2(a), red). Such porous film possessed an anomalous inverted refractive index in the visible, while in the near-infrared region, the absorption got higher (Fig. 2(b), red). This could significantly lower the performance of a possible nanophotonic device. When examining the relevant XRD pattern (Fig. 2(c), red), we can observe an increase of the (012) Sb peak at 29.3°, at the expense of the characteristic peaks of $Sb_2S_3$ at 15.7°, 17.7° and 25.1°. This suggests a loss of sulfur and the stoichiometric degradation of the film. Another confirmation of the sulfur loss comes from the Raman spectroscopy (see Fig. 2(d)), where we observed a significant increase of Raman peaks at 114 cm$^{-1}$ and 151 cm$^{-1}$, characteristic of Sb-Sb bonding in antimony and its oxide [30], at the expense of the stoichiometric $Sb_2S_3$ Raman peaks. These results, together with EDS maps (see Fig. S11), indicate that the porosity was caused by sulfur, which most probably stayed in the chamber after pre-ablation, formed small clusters on the silicon substrate, and later sublimated during the crystallization process while disrupting the whole film. Therefore, we suggest that the film's stoichiometric and optical quality do not depend on the pre-deposition pressure itself but rather on the time that samples spend in the deposition chamber after the pre-ablation. The scenario can be that the deposition starts immediately after the insertion of the sample so the sulfur cannot contaminate the sample, or the sample must be inserted into the chamber after the residual sulfur is pumped down and the pre-deposition pressure is reduced.

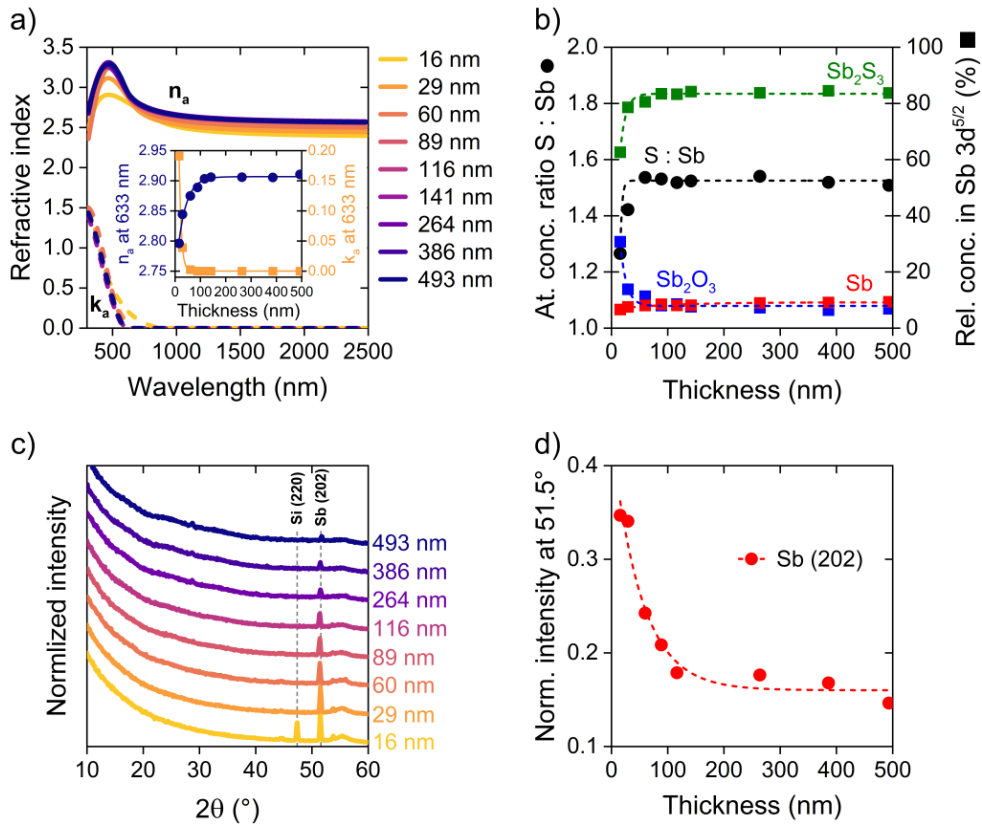

Fig. 3. A) Real (n) and imaginary (k) parts of the measured refractive index of amorphous $Sb_2S_3$ thin films with thicknesses listed in the graph as a function of the wavelength. The inset represents the refractive index and absorption coefficient of amorphous $Sb_2S_3$ films at 633 nm wavelength as functions of the film thickness. B) Atomic concentration ratio of S and Sb determined from the S $2p^{3/2}$ and Sb $3d^{5/2}$ peaks of fitted XPS spectra (black circle symbol), and the relative concentration of each antimony component (listed in the graph) determined from the Sb $3d^{5/2}$ XPS peak as functions of the film thickness (square symbols). C) Normalized XRD pattern of amorphous $Sb_2S_3$ thin films with thicknesses listed in the graph. The most characteristic peaks of Si and Sb are highlighted. D) Normalized XRD intensity at 51.5° diffraction angle that corresponds to the (202) plane of Sb crystals as a function of amorphous $Sb_2S_3$ film thickness. A-d) All scatter data are fitted by the exponential decay function with an offset to guide the eye.

## 4. Thickness dependence

A full understanding of the PLD of $Sb_2S_3$ gives the nanophotonic community another tool to fabricate a high-quality phase-change material with a significant optical contrast in the visible region. However, various nanophotonic applications require various thicknesses of this material. To verify if we can also use the optimized deposition parameters for films of different thicknesses, we inspected films with thicknesses ranging from 16 nm to 493 nm (Fig. S12(a)). The as-deposited amorphous films were again measured by XPS, capped with the $Al_2O_3$ protective film, and then characterized by spectroscopic ellipsometry and XRD. In Fig. 3(a), we can see that the refractive index increases, and the absorption coefficient decreases with the increasing thickness up to approx. 60 nm. Above 60 nm, the optical properties remain relatively independent of the thickness, as illustrated by the inset of Fig. 3(a).

To explain this behavior, we can analyze the corresponding XPS data: Namely, the S $2p^{3/2}$ and Sb $3d^{5/2}$ peak ratio (representing the overall stoichiometry, Fig. 3(b), left axis), and component analysis of the Sb $3d^{5/2}$ peak (given the mutual relative atomic concentration of $Sb_2S_3$, $Sb_2O_3$, and Sb components; Fig. 3(b), right axis). The low ratio indicates that there is a higher Sb content in thinner films. On the other hand, the high $Sb_2O_3$ component of these films suggests that the antimony becomes oxidized in the lower thicknesses. The facts are in line with an XRD measurement (Fig. 3(c)), where the Sb (202) peak at 51.5° is also exponentially increasing as the thickness goes down (Fig. 3(d)). The deposition of amorphous $Sb_2S_3$ films thus seems to be pretty robust and scalable for thicknesses above 60 nm. Thinner films, unfortunately, need further optimization, and care should be taken when using extrapolated deposition parameters obtained from thicker films.

The set of amorphous films was subsequently crystallized (27 °C/min) on a hotplate, again at 300 °C for 1 min. The thickness of each film decreased by approximately 12% during the crystallization (Fig. S12(a)), as expected [10]. For easier comparison, we will keep labeling the samples using amorphous thicknesses. Optical microscopy images in Fig. 4(a) reveal that the surface of the thinnest films is sometimes disrupted by many cracks. This unwanted effect is probably related to the sulfur contamination after the pre-ablation, as discussed above. In Fig. 4(b), we can see that the film disruption and the increased antimony content are responsible for the lowered index contrast at 633 nm in the thinnest films ($\Delta n \approx 0.4$). This is connected to a decrease of the absorption coefficient at 633 nm (Fig. 4(c)) from approx. 0.8 to only 0.1. However, the thicker films beyond 60 nm reached an impressive value of $\Delta n \approx 1.2$, which surpasses the average values mentioned above [16–21]. Although this increase is associated with an increased absorption, the very large contrast could be still beneficial in future applications beyond phase-only operation in the visible range. Note that a trend similar to that reported in Fig. 4(b) can be also observed at telecom wavelengths (Fig. S12(b)).

As all our films (except the thinnest ones) in the amorphous phase had almost identical optical properties, we ascribe the observed differences related to the process of their crystallization. We hypothesize that the large $\Delta n$ of polycrystalline films with thicknesses around 100 nm can be associated with a predominance of specific types of grain orientations. This connection has been established previously using micro ellipsometry, which confirmed that different $Sb_2S_3$ crystal planes have different refractive indices [14,18]. A careful XRD inspection of our sample series (Fig. 4(d)) fully supports that notion: In the films exhibiting the highest $\Delta n$, the peak at 51.1° increases at the expense of the peak at 51.5°. That corresponds to increasing the (212)/(260) crystal orientation of $Sb_2S_3$ at the expense of the (202) crystals. To better visualize the effect, the extracted normalized XRD intensity at 51.1° is plotted as a function of the film thickness in Fig. 4(e), strongly correlating with the $\Delta n$ progress in Fig. 4(a). We, therefore, argue that the substantial increase of the refractive index in the crystalline phase is associated with the presence (212)/(260) crystals of $Sb_2S_3$. A less pronounced but similar trend can also be seen for (111), (211), (220), and (531) orientations of $Sb_2S_3$ in Fig. S22.

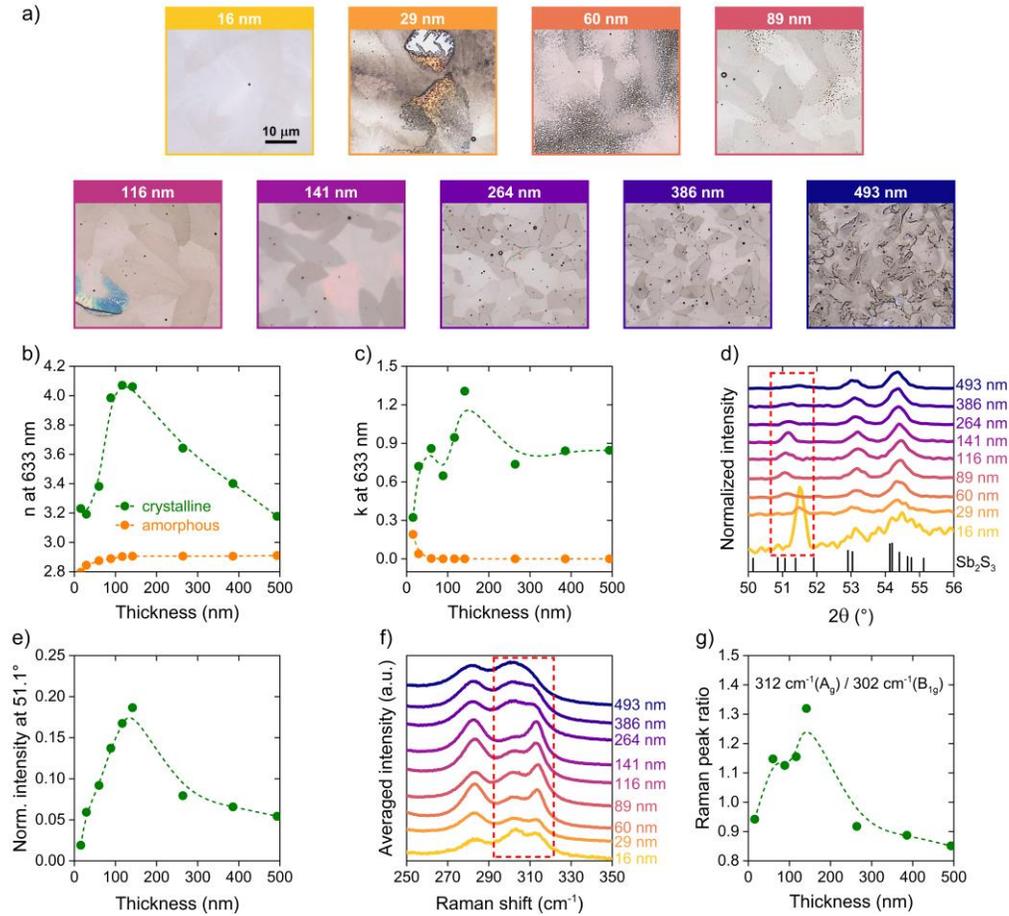

Fig. 4. a) Optical microscopy images of crystalline $Sb_2S_3$ films, deposited and crystallized using the previously optimized parameter set, with thicknesses listed in the images. b) Refractive index and (c) absorption coefficient of amorphous and crystalline $Sb_2S_3$ films at 633 nm wavelength as functions of the film thickness. d) Normalized XRD pattern and (f) averaged Raman spectra of crystalline $Sb_2S_3$ films with thicknesses listed in the graphs. e) Normalized XRD intensity at 51.1° diffraction angle in the area highlighted in (d) as a function of the film thickness. g) Intensity ratio between the two Raman peaks highlighted in (f) and listed in the graph as a function of the film thickness. b-g) All scatter data are fitted by a b-spline to guide the eye.

A similar correlation can be observed also in Raman spectra of the same sample series (Fig. 4(f)). Even though this measurement inherently averages over multiple crystal grains (see Methods), we can observe that the films with larger Δn systematically exhibit a larger ratio between the peaks at 312 cm$^{-1}$ and 302 cm$^{-1}$ (Fig. 4(g)). This ratio between the symmetric ($A_g$) and anti-symmetric ($B_{1g}$) stretching modes of the Sb-S bond [31] can thus be considered as another indicator of the largest refractive index and, therefore, the largest index contrast during the crystallization. A similar connection has been proposed recently [8] when the large optical contrast was ascribed to the ratio between 303 cm$^{-1}$ ($A_g$) and 281 cm$^{-1}$ ($A_g$) Raman peaks being close to 1.5 [8]. Our results partially contradict this claim, as this ratio was below 1 for our films with the largest contrast. We hypothesize that the ratio in [8] was set for the bulk $Sb_2S_3$ and might vary for the deposited films. These contradictory results prove that the interplay between the structural and optical properties surely requires a more thorough investigation of individual crystal grains.

## 5. Conclusions

In conclusion, we optimized the PLD of $Sb_2S_3$ films with varying thicknesses for their use in tunable nanophotonics. While $Sb_2S_3$ films deposited by PLD have been studied before, the optimization of this deposition technique focusing on the contrast of optical properties during the $Sb_2S_3$ transition has been missing. Moreover, the thickness scalability of this technique has not been discussed as well. We have investigated the influence of various deposition and crystallization parameters of PLD on the physical properties of phase-changing $Sb_2S_3$ films. We have proved that the ideal $Sb_2S_3$ stoichiometry relates to a very low absorption in the amorphous phase (0.03 at 633 nm wavelength). Such absorption is close to the average literature value of 0.02 and introduces a tunable material that could compete with other low-absorbing non-tunable dielectric materials. By gradual thermal crystallization, we have improved the integrity of the crystallized film and increased the refractive index contrast up to 1.2 at 633 nm wavelength. This value, being 1.5 larger than the average contrast values reported so far, indicates that careful optimization can significantly improve tuning and switching capacity of nanophotonic devices in the visible. Moreover, the low absorption and large refractive index contrast in the near-infrared have proved that films deposited by PLD are also very promising for applications in this spectral range. Then, we tested this optimized procedure for various film thicknesses. By first looking at the optical and structural properties of amorphous films, we have found that contamination by residual sulfur after the pre-ablation might play a key detrimental role in the deposition of very thin films. When looking at the polycrystalline phase of these films, we found that there is slightly different preferential grain orientation for different thicknesses. As $Sb_2S_3$ crystals are generally anisotropic and refractive indices depend on crystal grain orientations, we have observed a strong dependence of the refractive index and its contrast on the film thickness. We ascribe the largest contrast to the prevalence of some specific crystal orientations ((212)/(260)), reflected also in the large ratio of the corresponding Raman peaks (312 cm$^{-1}$ / 302 cm$^{-1}$). Our work provides the most optimal deposition parameters and describes their connection with particular optical qualities of the films. It also highlights stoichiometric and crystallographic engineering as powerful tools in the development of high-quality phase-changing materials for nanophotonics. We have also identified that a detailed investigation of individual crystal grains and especially the ways how to control their formation could lead to even further improved nanophotonic devices with even better tunability.

**Acknowledgments.** This work is supported by the Grant Agency of the Czech Republic (21-29468S), the European Commission (H2020 – Twinning project no. 810626 – SINNCE, M-ERA NET HYSUCAP/TACR-TH71020004), Brno Ph.D. talent scholarship, Thermo Fisher Scientific scholarship, and Quality Internal Grants of BUT (KInG BUT), Reg. No. CZ.02.2.69/0.0/0.0/19_073/0016948, which is financed from the OP RDE. We also acknowledge CzechNanoLab Research Infrastructure supported by MEYS CR (LM2018110) and thank Dr. Jan Mistrik for his generous help with the ellipsometry measurements.

**Disclosures.** The authors declare no conflicts of interest.

## References

1. M. Wuttig and N. Yamada, "Phase-change materials for rewriteable data storage," Nature Mater **6**, 824–832 (2007).
2. S. Abdollahramezani, O. Hemmatyar, H. Taghinejad, A. Krasnok, Y. Kiarashinejad, M. Zandehshahvar, A. Alù, and A. Adibi, "Tunable nanophotonics enabled by chalcogenide phase-change materials," Nanophotonics **9**(5), 1189-1241 (2020).
3. S. Cueff, J. John, Z. Zhang, J. Parra, J. Sun, R. Orobtchouk, S. Ramanathan, and P. Sanchis, "VO2 nanophotonics," APL Photonics **5**, 110901 (2020).
4. P. Kepič, F. Ligmajer, M. Hrtoň, H. Ren, L. S. Menezes, S. A. Maier, and T. Šikola, "Optically Tunable Mie Resonance VO2 Nanoantennas for Metasurfaces in the Visible," ACS Photonics **8**(4), 1048-1057 (2021).
5. K. Shportko, S. Kremers, M. Woda, D. Lencer, J. Robertson, and M. Wuttig, "Resonant bonding in crystalline phase-change materials," Nature Mater **7**, 653–658 (2008).
6. L. M. Monier, C. C. Popescu, L. Ranno, B. Mills, S. Geiger, D. Callahan, M. Moebius, and J. Hu, "Endurance of chalcogenide optical phase change materials: a review," Opt. Mater. Express **12**, 2145-2167 (2022).

7. W. Dong, H. Liu, J. K. Behera, L. Lu, R. J. H. Ng, K. V. Sreekanth, X. Zhou, J. K. W. Yang, and R. E. Simpson, "Wide Bandgap Phase Change Material Tuned Visible Photonics," Adv. Funct. Mater **29**, 1806181 (2019).
8. Y. Gutierrez, A. P. Ovvyan, G. Santos, D. Juan, S. A. Rosales, J. Junquera, P. García-Fernández, S. Dicorato, M. Giangregorio, E. Dilonardo, F. Palumbo, M. Modreanu, J. Resl, O. Ischenko, G. Garry, T. Jouzi, M. Gheorghe, C. Cobianu, K. Hingerl, C. Cobet, F. Moreno, W. H. Pernice, and M. Losurdo, "Interlaboratory Study on Sb2S3 Interplay between Structure, Dielectric Function and Amorphous-to-Crystalline Phase Change for Photonics," iScience **25**(6), 104377 (2022).
9. G. Mirabelli, C. McGeough, M. Schmidt, E. K. McCarthy, S. Monaghan, I. M. Povey, M. McCarthy, F. Gity, R. Nagle, G. Hughes, A. Cafolla, P. K. Hurley, and R. Duffy, "Air sensitivity of MoS2, MoSe2, MoTe2, HfS2, and HfSe2," J. Appl. Phys. 120, 125102 (2016).
10. M. Delaney, I. Zeimpekis, D. Lawson, D. W. Hewak, and O. L. Muskens, "A New Family of Ultralow Loss Reversible Phase-Change Materials for Photonic Integrated Circuits: Sb2S3 and Sb2Se3," Adv. Funct. Mater. **30**, 2002447 (2020).
11. K. Gao, K. Du, S. Tian, H. Wang, L. Zhang, Y. Guo, B. Luo, W. Zhang, and T. Mei, "Intermediate Phase-Change States with Improved Cycling Durability of Sb2S3 by Femtosecond Multi-Pulse Laser Irradiation," Adv. Funct. Mater. **31**, 2103327 (2021).
12. P. Arun and A.G. Vedeshwar, "Effect of heat treatment on the optical properties of amorphous Sb2S3 film: The possibility of optical storage," Journal of Non-Crystalline Solids **220**(1), 63-68 (1997).
13. Y. Gutiérrez, A. Fernández-Pérez, S. A. Rosales, C. Cobianu, M. Gheorghe, M. Modreanu, J. M. Saiz, F. Moreno, and M. Losurdo, "Polarimetry analysis and optical contrast of Sb2S3 phase change material," Opt. Mater. Express **12**, 1531-1541 (2022).
14. M. Schubert, T. Hofmann, C. M. Herzinger, and W. Dollase, "Generalized ellipsometry for orthorhombic, absorbing materials: dielectric functions, phonon modes and band-to-band transitions of Sb2S3," Thin Solid Films 455, 619–623 (2004).
15. M. Schubert and W. Dollase, "Generalized ellipsometry for biaxial absorbing materials: determination of crystal orientation and optical constants of Sb2S3," Opt. Lett. 27, 2073 (2002).
16. Z. Fang, J. Zheng, A. Saxena, J. Whitehead, Y. Chen, and A. Majumdar, "Non-Volatile Reconfigurable Integrated Photonics Enabled by Broadband Low-Loss Phase Change Material," Adv. Optical Mater **9**, 2002049 (2021).
17. L. Lu, Z. Dong, F. Tijiptoharsono, R. J. H. Ng, H. Wang, S. D. Rezaei, Y. Wang, H. S. Leong, P. C. Lim, J. K. W. Yang, and Robert E. Simpson, "Reversible Tuning of Mie Resonances in the Visible Spectrum," ACS Nano **15**(12), 19722-19732 (2021).
18. H. Liu, W. Dong, H. Wang, L. Lu, Q. Ruan, Y. S. Tan, R. E. Simpson, and J. K. W. Yang, "Rewritable color nanoprints in antimony trisulfide films," Sci. Adv. **6**, 7171–7187 (2020).
19. M. Delaney, I. Zeimpekis, H. Du, X. Yan, M. Banakar, D. J. Thomson, D. W. Hewak, and O. L. Muskens, "Nonvolatile programmable silicon photonics using an ultralow-loss Sb2Se3 phase change material," Sci. Adv. **7**(25), (2021).
20. T. Y. Teo, M. Krbal, J. Mistrik, J. Prikryl, L. Lu, and R. E. Simpson, "Comparison and analysis of phase change materials-based reconfigurable silicon photonic directional couplers," Opt. Mater. Express **12**, 606-621 (2022).
21. P. Moitra, Y. Wang, X. Liang, L. Lu, A. Poh, T. W.W. Mass, R. E. Simpson, A. I. Kuznetsov, and R. Paniagua-Dominguez, "Tunable wavefront control in the visible spectrum using low-loss chalcogenide phase change metasurfaces," https://arxiv.org/abs/2206.07628.
22. S. Mahanty, J. M. Merino, and M. León, "Preparation and optical studies on flash evaporated Sb2S3 thin films," Journal of Vacuum Science & Technology A **15**, 3060 (1997).
23. C. Cobianu, M. Gheorghe, M. Modreanu, Y. Gutierrez, and M. Losurdo, "Chemically bath deposited Sb2S3 films as optical phase change materials," in 2021 International Semiconductor Conference (CAS) (IEEE, 2021), pp. 249–252.
24. D. H. Kim, S. J. Lee, M. S. Park, J. K. Kang, J. H. Heo, S. H. Im, and S. J. Sung, "Highly reproducible planar Sb2S3-sensitized solar cells based on atomic layer deposition," Nanoscale **6**, 14549-14554 (2014).
25. D. Santos-Cruz, M. de la L. Olvera-Amador, S. A. Mayen-Hernandez, J. G. Quiñones-Galván, J. Santos-Cruz, and F. de Moure-Flores, "Structural, optical, and morphological characterization of Sb2S3 thin films grown by pulsed laser deposition," Journal of Laser Applications **33**, 042012 (2021).
26. I. S. Virt, I. O. Rudyj, I. V. Kurilo, I. Ye. Lopatynskyi, L. F. Linnik, V. V. Tetyorkin, P. Potera, and G. Luka, "Properties of Sb2S3 and Sb2Se3 thin films obtained by pulsed laser ablation," Semiconductors **47**, 1003–1007 (2013).
27. D. Nečas and P. Klapetek, "Gwyddion: an open-source software for SPM data analysis," Cent. Eur. J. Phys. **10**(1), 181-188 (2012).
28. P. Arun and A. G. Vedeshwar, "Phase modification by instantaneous heat treatment of Sb2S3 films and their potential for photothermal optical recording," Journal of Applied Physics **79**, 4029-4036 (1996).
29. M. S. Dawood, A. Hamdana, and J. Margota, "Influence of surrounding gas, composition and pressure on plasma plume dynamics of nanosecond pulsed laser-induced aluminum plasmas," AIP Advances **5**, 107143 (2015).
30. R. Parize, T. Cossuet, O. Chaix-Pluchery, H. Roussel, E. Appert, and V. Consonni, "In situ analysis of the crystallization process of Sb2S3 thin films by Raman scattering and X-ray diffraction," Mater. Des. **121**, 1–10 (2017).

31. S. Kharbish, E. Libowitzky, and A. Beran, "Raman spectra of isolated and interconnected pyramidal XS3 groups (X = Sb, Bi) in stibnite, bismuthinite, kermesite, stephanite and bournonite," European Journal of Mineralogy **21**(2), 325 – 333 (2009).

# Pulsed laser deposition of $Sb_2S_3$ films for phase-change tunable nanophotonics: supplemental document

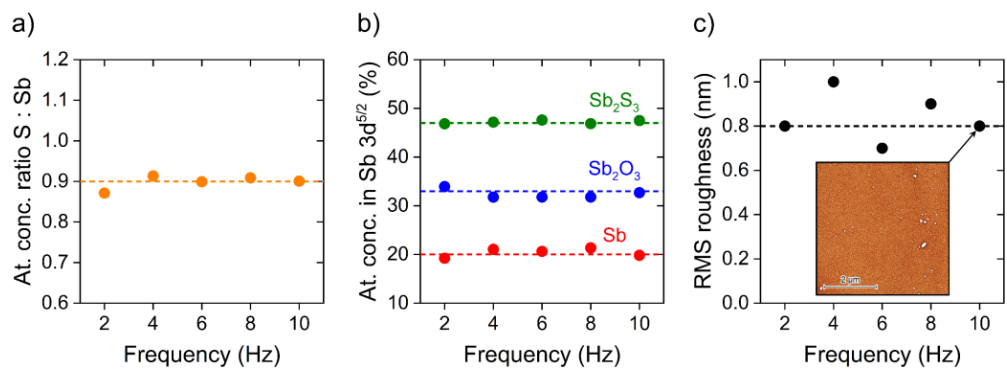

Fig. S1. Optimization of laser pulse frequency by measuring samples deposited at 1.5 J/cm$^2$, 30 mTorr, and 5000 pulses (approx. 160 nm thick). a) Concentration ratio of S and Sb atoms determined from S 2p$^{3/2}$ and Sb 3d$^{5/2}$ peak components of fitted XPS spectra (see Fig. S3) as a function of the deposition laser pulse frequency. The dashed line denotes an average value of 0.9. b) Atomic concentration of various film components (listed in the graph) determined from the Sb 3d$^{5/2}$ XPS peak as functions of the laser pulse frequency. The dashed lines denote the average value of each component. c) RMS roughness, extracted from 5 × 5 µm$^2$ AFM images (such as the inset for the 10 Hz sample) as a function of the laser pulse frequency.

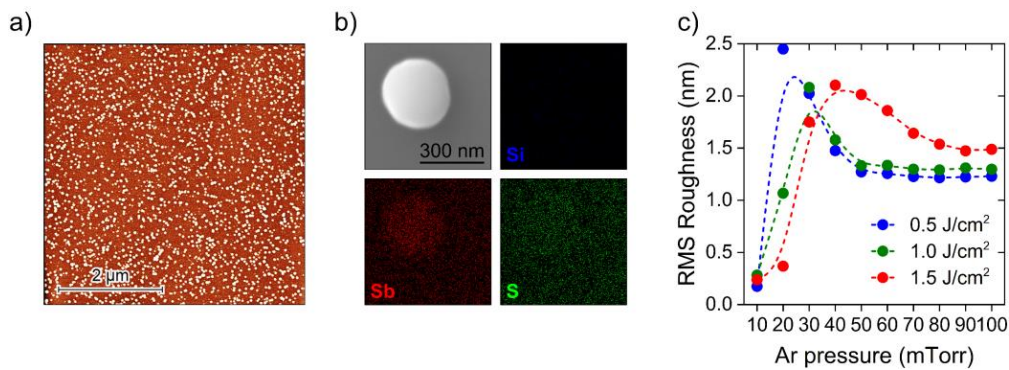

Fig. S2. a) AFM image of the $Sb_2S_3$ film deposited at 10 Hz, 0.5 J/cm$^2$, 20 mTorr, and 5000 pulses (approx. 130 nm thick). b) SEM micrograph of a typical crystal on the top of the sample in (a) with the corresponding EDS maps confirming that the crystals are formed mainly by pure antimony. c) RMS roughness of $Sb_2S_3$ films deposited at 10 Hz, 5000 pulses, and laser fluences listed in the graph as functions of the deposition Ar pressure. The data were obtained from 5×5 µm$^2$ AFM images and fitted by a b-spline.

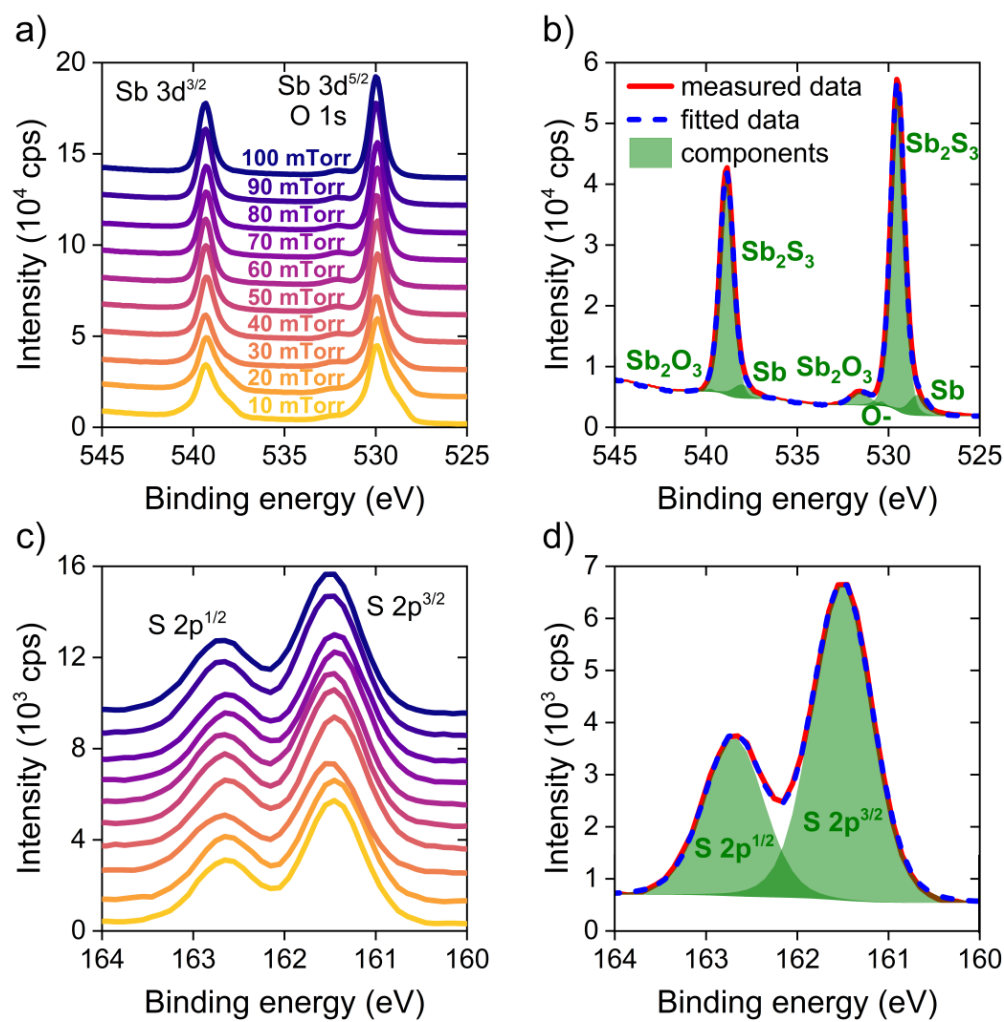

Fig. S3. a) Sb 3d (doublet) and O 1s peak regions of the measured XPS spectra of samples deposited at 10 Hz, 0.5 J/cm$^2$, 5000 pulses and various Ar pressures listed in the graph. b) The fitting process is done according to [1] for the sample deposited at an Ar pressure of 100 mTorr: First, the whole spectrum is shifted according to a C 1s (284.8 eV) reference peak. Second, as the O 1s peak overlaps with the Sb 3d$^{5/2}$ sub-peak, we independently fit the Sb 3d3/2 region with Sb2O3, Sb2S3 and Sb components. Third, the corresponding components of the Sb 3d$^{5/2}$ are constrained by setting the same full width at half maximum (FWHM) and the area ratio 2:3 between the doublet components. Fourth, we add the O 1s sub-peaks corresponding to Sb bonding (O-Sb) and adventitious species associated with carbon and hydrogen and fit components of the s. In case the Sb$_2$O$_3$ and Sb components exhibit overlapping positions with Sb$_2$S$_3$ or FWHM larger than 3 eV, position constraints are added based on the references in Tab. S1. In Tab. S1, we present the positions of all peaks for each sample compared to the reference of the NIST XPS Database [2]. c) S 2p peaks of the measured XPS spectra of the samples deposited at 10 Hz, 0.5 J/cm$^2$, 5000 pulses and various Ar pressures listed in (a). d) The process of fitting S 2p XPS regions is explained for the sample deposited at 100 mTorr of Ar: First, the whole spectrum is shifted according to a C 1s (284.8 eV) reference peak. Second, the doublet components are constrained by equal FWHM and 1:2 area ratio and fitted.

| Fluence (J/cm²) | Pressure (mTorr) | Sb 3d³ᐟ² | | | Sb 3d⁵ᐟ² | | | O 1s | | S 2p¹ᐟ² | S 2p³ᐟ² |
|---|---|---|---|---|---|---|---|---|---|---|---|
| | | Sb₂O₃ | Sb₂S₃ | Sb | Sb₂O₃ | Sb₂S₃ | Sb | O-Sb | O- | | |
| 0.5 | 10 | 539.8 | 538.9 | 537.8 | 530.4 | 529.5 | 528.4 | 530.2 | 531.4 | 162.7 | 161.5 |
| | 20 | 539.9 | 538.9 | 537.8 | 530.4 | 529.5 | 528.4 | 529.8 | 531.4 | 162.7 | 161.5 |
| | 30 | 539.6 | 538.9 | 537.9 | 530.4 | 529.5 | 528.4 | 530.2 | 531.9 | 162.7 | 161.5 |
| | 40 | 539.6 | 538.9 | 537.9 | 530.4 | 529.5 | 528.4 | 530.2 | 531.7 | 162.7 | 161.5 |
| | 50 | 539.6 | 538.9 | 537.9 | 530.4 | 529.5 | 528.4 | 530.2 | 531.7 | 162.7 | 161.5 |
| | 60 | 539.7 | 538.8 | 537.9 | 530.4 | 529.5 | 528.4 | 530.2 | 531.6 | 162.7 | 161.5 |
| | 70 | 539.9 | 538.9 | 538.0 | 530.4 | 529.5 | 528.4 | 530.2 | 531.6 | 162.7 | 161.5 |
| | 80 | 539.8 | 538.9 | 538.0 | 530.4 | 529.5 | 528.4 | 530.2 | 531.6 | 162.7 | 161.5 |
| | 90 | 539.8 | 538.8 | 538.0 | 530.4 | 529.5 | 528.4 | 530.2 | 531.5 | 162.7 | 161.5 |
| | 100 | 539.8 | 538.9 | 538.1 | 530.4 | 529.5 | 528.4 | 530.2 | 531.6 | 162.7 | 161.5 |
| 1.0 | 10 | 539.9 | 538.9 | 537.9 | 530.4 | 529.5 | 528.4 | 530.2 | 531.0 | 162.6 | 161.4 |
| | 20 | 539.9 | 538.9 | 537.9 | 530.4 | 529.5 | 528.4 | 530.2 | 531.1 | 162.6 | 161.4 |
| | 30 | 539.8 | 538.9 | 537.9 | 530.4 | 529.5 | 528.4 | 530.2 | 531.1 | 162.7 | 161.5 |
| | 40 | 539.8 | 538.9 | 537.9 | 530.4 | 529.5 | 528.4 | 530.2 | 531.3 | 162.7 | 161.5 |
| | 50 | 539.6 | 538.9 | 537.9 | 530.4 | 529.5 | 528.4 | 530.2 | 531.6 | 162.6 | 161.4 |
| | 60 | 539.9 | 538.9 | 538.0 | 530.4 | 529.5 | 528.4 | 530.2 | 531.3 | 162.6 | 161.4 |
| | 70 | 539.8 | 538.9 | 538.0 | 530.4 | 529.5 | 528.4 | 530.2 | 531.2 | 162.6 | 161.4 |
| | 80 | 539.9 | 538.9 | 538.1 | 530.4 | 529.5 | 528.4 | 529.8 | 531.2 | 162.6 | 161.5 |
| | 90 | 539.8 | 538.9 | 538.0 | 530.4 | 529.5 | 528.4 | 530.2 | 531.3 | 162.6 | 161.4 |
| | 100 | 539.8 | 538.9 | 538.0 | 530.4 | 529.5 | 528.4 | 530.2 | 531.2 | 162.6 | 161.4 |
| 1.5 | 10 | 539.9 | 538.9 | 537.8 | 530.4 | 529.5 | 528.4 | 531.1 | 530.1 | 162.6 | 161.4 |
| | 20 | 539.9 | 538.9 | 537.9 | 530.4 | 529.5 | 528.4 | 531.1 | 530.1 | 162.4 | 161.2 |
| | 30 | 539.6 | 538.9 | 537.8 | 530.4 | 529.5 | 528.4 | 531.4 | 530.2 | 162.7 | 161.5 |
| | 40 | 539.6 | 538.8 | 537.9 | 530.4 | 529.5 | 528.4 | 531.6 | 530.2 | 162.7 | 161.5 |
| | 50 | 539.9 | 538.9 | 538.0 | 530.4 | 529.5 | 528.6 | 531.4 | 530.2 | 162.7 | 161.5 |
| | 60 | 539.7 | 538.9 | 538.2 | 530.4 | 529.5 | 528.8 | 531.4 | 529.8 | 162.7 | 161.5 |
| | 70 | 539.8 | 538.9 | 538.0 | 530.4 | 529.5 | 528.4 | 531.3 | 530.2 | 162.7 | 161.5 |
| | 80 | 540.0 | 538.9 | 538.0 | 530.4 | 529.5 | 528.4 | 531.5 | 530.2 | 162.6 | 161.4 |
| | 90 | 539.9 | 538.9 | 538.0 | 530.4 | 529.5 | 528.4 | 531.4 | 530.2 | 162.6 | 161.4 |
| | 100 | 539.9 | 538.9 | 538.0 | 530.4 | 529.5 | 528.4 | 531.4 | 530.2 | 162.6 | 161.4 |
| Reference[2] | | 539.8 ± 0.2 | 539.2 ± 0.1 | 538 ± 0.2 | 530.1 ± 0.3 | 529.5 ± 0.5 | 528.2 ± 0.2 | 530 ± 0.3 | - | 162.3 ± 0.3 | 161.3 ± 0.3 |

Tab. S1. The binding energy of each fitted XPS component in Sb 3d and S 2p regions of the samples deposited at 10 Hz, 5000 pulses, at the listed fluences and Ar pressures. Fitted positions are compared to the averaged positions in the NIST XPS Database [2].

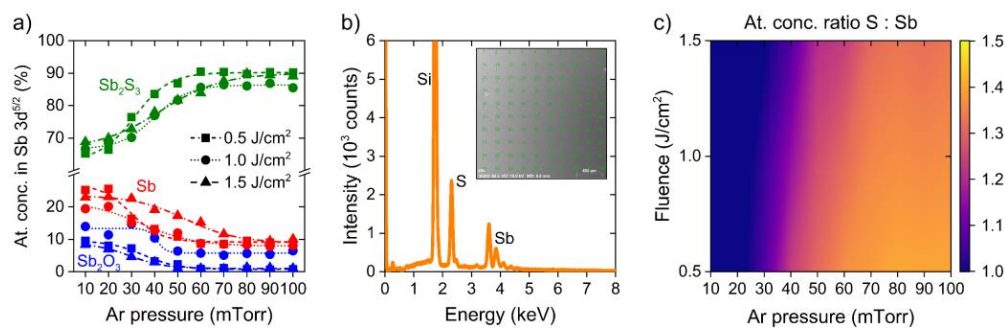

Fig. S4. a) Atomic concentration of various components (listed in the graph) obtained by fitting the Sb $3d^{5/2}$ XPS peak as a function of the deposition Ar pressure. The data were fitted by a sigmoidal Boltzmann function. b) EDS spectrum of an $Sb_2S_3$ film on a Si substrate deposited at 10 Hz, 0.5 J/cm$^2$, 100 mTorr, and 5000 pulses. The inset illustrates that the spectrum was acquired by averaging 100 spots, with a net exposure time of 2 min. c) Atomic concentration ratio between S and Sb as a 2D map of the deposition laser fluence and Ar pressure. The ratio was obtained from the fitted EDS spectrum (using the PB-ZAF method), averaging 100 spots in three locations on each sample. Note that the atomic concentration ratio in the EDS map is lower than in the XPS map in Fig. 2 by 0.1 due to the dependence of the EDS method on the electron and X-ray penetration depth. However, the trend is qualitatively the same.

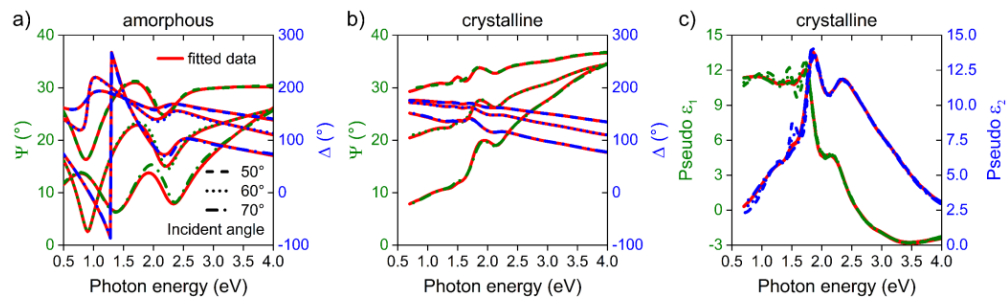

Fig. S5. Measured and fitted ellipsometry spectra (psi and delta) of (a) amorphous and (b) crystalline $Sb_2S_3$ film, deposited at 10 Hz, 0.5 J/cm$^2$, 100 mTorr and 5000 pulses and annealed for 2 minutes at 300 °C by 27 °C/min ramp rate. c) Measured and fitted pseudo-dielectric function of the crystalline $Sb_2S_3$ as a function of photon energy. The model of the Si substrate and three layers was used, where the thicknesses of the native $SiO_2$ (ca. 2 nm) layer and $Al_2O_3$ protective layer (ca. 15 nm) were determined by the reference measurements. The $Sb_2S_3$ layer was fitted by one Tauc-Lorentz oscillator in (a) and five Tauc-Lorentz oscillators in (b-c). The five oscillators in the crystalline phase were chosen based on the polycrystalline character of the anisotropic $Sb_2S_3$ film [3] and an initial convergence test (Fig. S6). Parameters of the fitted oscillators for this specific sample can be seen in Tab. S3. Most of the films from the optimization series were fitted using the same number of oscillators, but some films required only 4 or 3 oscillators in their crystallized phase. For clarity, we present only the fitted real (n) and imaginary (k) parts of the refractive indices (see Fig. S7) of all measured films. The fitted data or oscillator parameters of all samples from Fig. 1 can be provided upon a reasonable request.

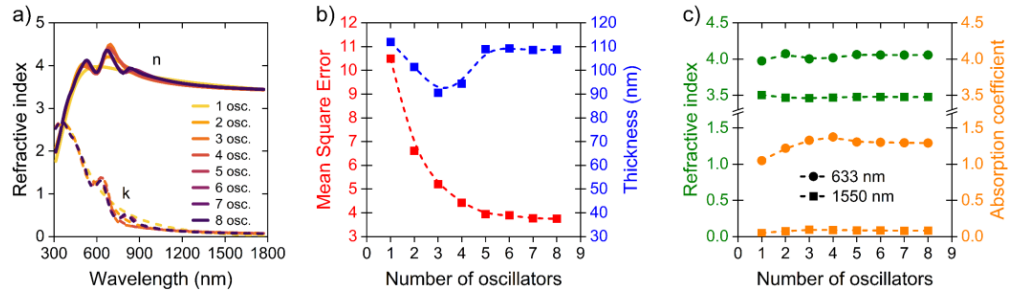

Fig. S6. Convergence test of a number of Tauc-Lorentz oscillators in the ellipsometry fitting of the crystalline $Sb_2S_3$ film deposited at 10 Hz, 0.5 J/cm$^2$, 100 mTorr and 5000 pulses (**141 nm**) and annealed at 300°C by 27°C/min ramp rate. a) Real (n) and imaginary (k) parts of the measured refractive index as functions of the wavelength for various numbers of Tauc-Lorentz oscillators used for the ellipsometry fitting. b) Mean square error of the ellipsometry fitting and the fitted thickness as a function of the number of oscillators used. c) The refractive index and the absorption coefficient at 633 nm and 1550 nm wavelengths as functions of the number of oscillators used. The refractive index, absorption, and thickness do not significantly change after 5 oscillators. The convergence test was done to investigate the difference between refractive indices (often) modeled by one Tauc-Lorentz oscillator [4-6] and multiple oscillators, which are physically more correct due to polycrystallinity and anisotropy of the sample [3]. b-c) Data are fitted by b-splines for clarity.

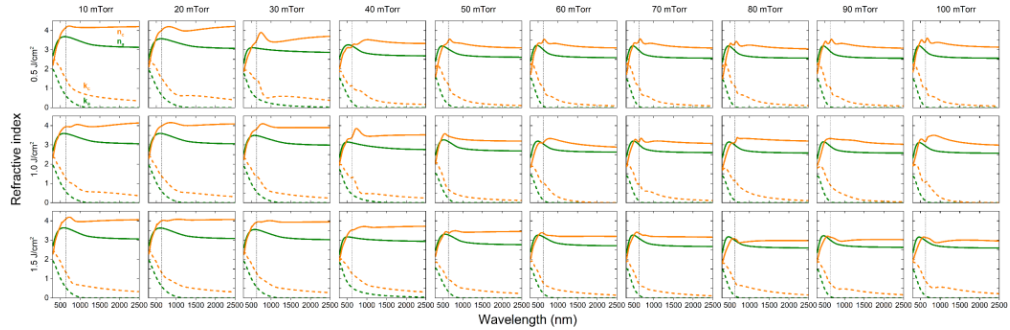

Fig. S7. Real (n, full line) and imaginary (k, dashed line) parts of the measured refractive index spectra of amorphous (green) and crystalline (orange) $Sb_2S_3$ films, deposited at 10 Hz, 5000 pulses and different values of laser fluence and Ar pressure listed in the graph. The data were obtained by fitting a theoretical model to spectroscopic ellipsometry measurements (as in Fig. S5). Black dashed lines highlight the investigated wavelength of 633 nm.

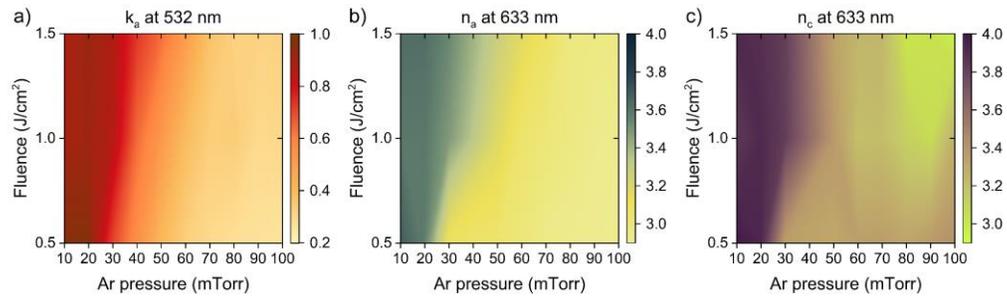

Fig. S8. Maps of the complex refractive index of $Sb_2S_3$ as 2D functions of the deposition laser fluence and Ar pressure. a) Absorption coefficient of amorphous $Sb_2S_3$ films at 532 nm wavelength. The refractive index of (b) amorphous and (c) crystalline $Sb_2S_3$ films at 633 nm wavelength.

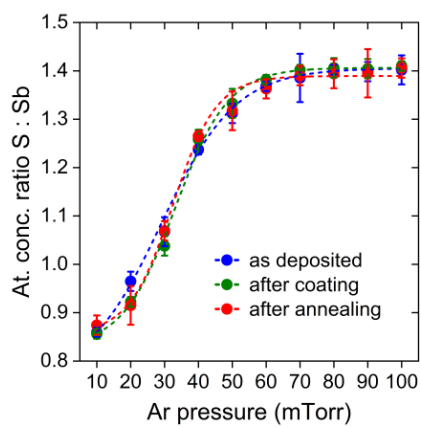

Fig. S9. EDS-estimated atomic concentration ratio between S and Sb of $Sb_2S_3$ films deposited at 10Hz, 0.5 J/cm$^2$ and 5000 pulses as a function of the deposition Ar pressure. The films were measured immediately after the deposition, coating, and crystallization (300 °C, 2 min, 160 °C/min, air). The error bars were calculated by measuring 3 different spots on each sample (as described in Fig. S4). The data are fitted by a sigmoidal Boltzmann function.

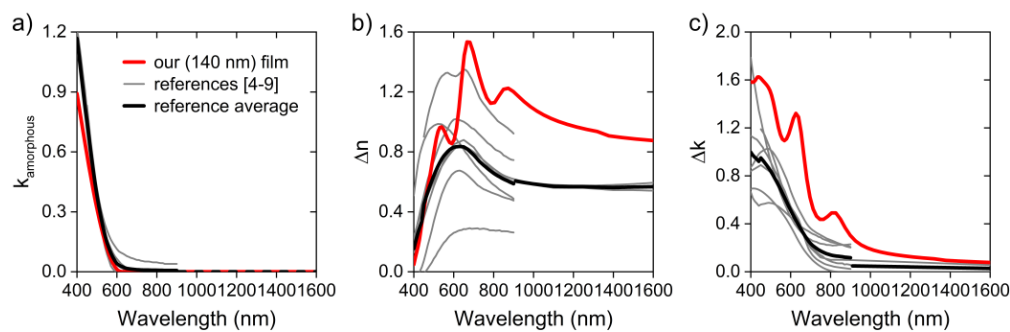

Fig. S10. Comparison of (a) absorption coefficients, (b) refractive index and (c) absorption contrasts of our 140 nm amorphous thin film, deposited at 10 Hz, 0.5 J/cm$^2$, 100 mTorr and 5000 pulses, with specific and averaged values of various $Sb_2S_3$ films from references [4-9].

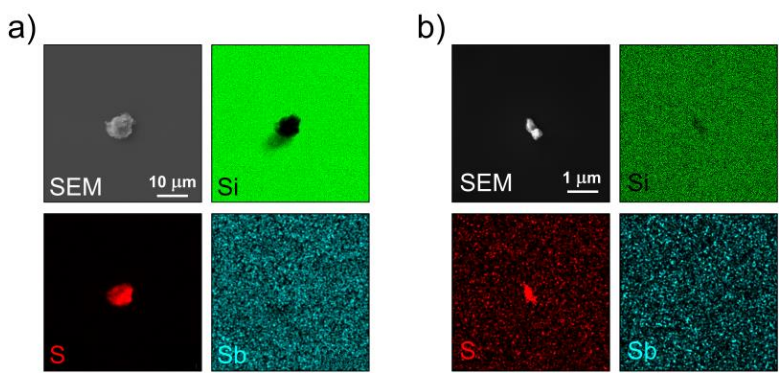

Fig. S11. a-b) SEM micrographs and EDS maps of a bare Si substrate, inserted into the deposition chamber after the target pre-ablation and left there overnight until the pressure dropped to 8e-5 mTorr. We focused on two different particles on top of the sample.

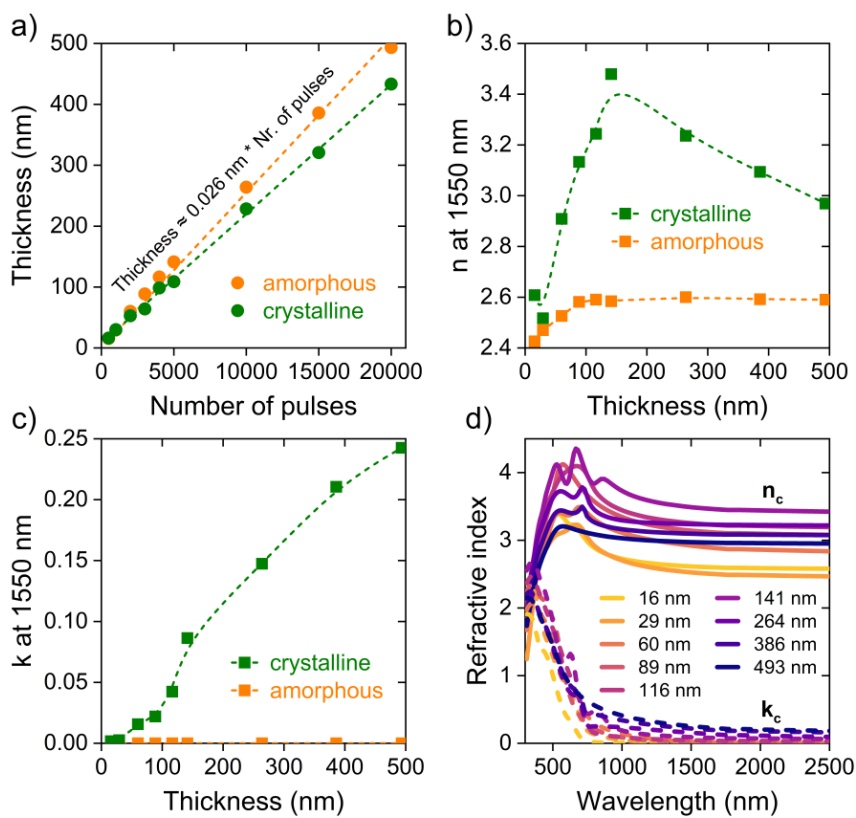

Fig. S12. a) Thickness of amorphous and crystalline $Sb_2S_3$ films, obtained from the profilometry measurement and confirmed by the spectroscopic ellipsometry, as a function of the number of deposition laser pulses. The data are fitted with linear functions. The inscribed equation represents an approximate formula relating the pulse number with the thickness of the amorphous film. b) Refractive index and (c) absorption coefficient of the amorphous and crystalline $Sb_2S_3$ films at 1550 nm wavelength as functions of the film thickness. The experimental data are fitted by a b-spline. d) Real (n) and imaginary (k) parts of the refractive index of the crystalline $Sb_2S_3$ films, with the thicknesses listed in the graph. We did the same convergence test for each thickness as in Fig. S6 (see Figs. S14-S21).

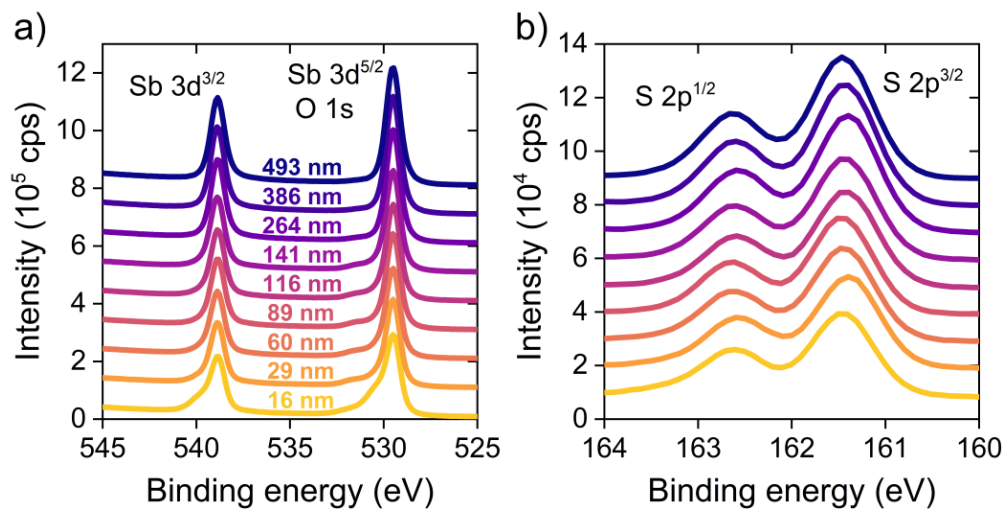

Fig. S13. a) Sb 3d, O 1s and (b) S 2p peaks of the measured XPS spectra of amorphous $Sb_2S_3$ films deposited at 10 Hz, 0.5 J/cm$^2$, 100 mTorr of Ar and with thicknesses listed in the graph.

| Pulses | Thickness (nm) | Sb 3d$^{3/2}$ | | | Sb 3d$^{5/2}$ | | | O 1s | | S 2p$^{1/2}$ | S 2p$^{3/2}$ |
|---|---|---|---|---|---|---|---|---|---|---|---|
| | | $Sb_2O_3$ | $Sb_2S_3$ | Sb | $Sb_2O_3$ | $Sb_2S_3$ | Sb | O-Sb | O- | | |
| 500 | 16 | 539.7 | 538.8 | 538.0 | 530.4 | 529.5 | 528.6 | 531.7 | 530.1 | 162.6 | 161.5 |
| 1000 | 29 | 539.6 | 538.9 | 538.1 | 530.4 | 529.5 | 528.6 | 531.6 | 530.1 | 162.6 | 161.4 |
| 2000 | 60 | 539.6 | 538.9 | 538.1 | 530.4 | 529.5 | 528.6 | 531.5 | 530.1 | 162.6 | 161.4 |
| 3000 | 89 | 539.6 | 538.9 | 538.1 | 530.4 | 529.5 | 528.6 | 531.5 | 530.2 | 162.7 | 161.5 |
| 4000 | 116 | 539.6 | 538.9 | 538.1 | 530.4 | 529.5 | 528.6 | 531.4 | 530.1 | 162.6 | 161.4 |
| 5000 | 141 | 539.6 | 538.9 | 538.2 | 530.4 | 529.5 | 528.6 | 531.4 | 530.1 | 162.6 | 161.4 |
| 10000 | 264 | 539.6 | 538.8 | 538.2 | 530.4 | 529.5 | 528.6 | 531.4 | 530.1 | 162.6 | 161.4 |
| 15000 | 386 | 539.2 | 538.9 | 538.1 | 530.1 | 529.5 | 528.6 | 531.3 | 530.1 | 162.6 | 161.4 |
| 20000 | 493 | 539.2 | 538.9 | 538.1 | 530.1 | 529.5 | 528.6 | 531.4 | 530.2 | 162.6 | 161.4 |
| Reference[2] | | 539.8 ± 0.2 | 539.2 ± 0.1 | 538 ± 0.2 | 530.1 ± 0.3 | 529.5 ± 0.5 | 528.2 ± 0.2 | 530 ± 0.3 | - | 162.3 ± 0.3 | 161.3 ± 0.3 |

Tab. S2. The binding energy of each fitted XPS component in Sb 3d and S 2p peaks of samples deposited at 10 Hz, 0.5 J/cm$^{-2}$, 100 mTorr and pulses/thicknesses listed in the graph. The fitted positions are compared to the averaged positions in the NIST XPS Database [2].

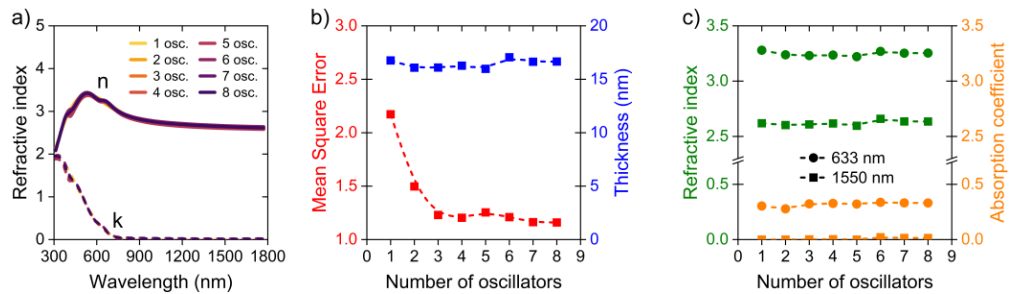

Fig. S14. Convergence test of a number of Tauc-Lorentz oscillators in the ellipsometry fitting of the crystalline $Sb_2S_3$ film deposited at 10 Hz, 0.5 J/cm$^2$, 100 mTorr and 500 pulses (**16 nm**) and annealed at 300°C by 27°C/min ramp rate. a) Real (n) and imaginary (k) parts of the measured refractive index as functions of the wavelength for various numbers of Tauc-Lorentz oscillators used for the ellipsometry fitting. b) Mean square error of the ellipsometry fitting and the fitted thickness as functions of the number of oscillators used. c) The refractive index and the absorption coefficient at 633 nm and 1550 nm wavelengths as functions of the number of oscillators used. The refractive index, absorption, and thickness do not significantly change after 3 oscillators. b-c) Data are fitted by a b-spline.

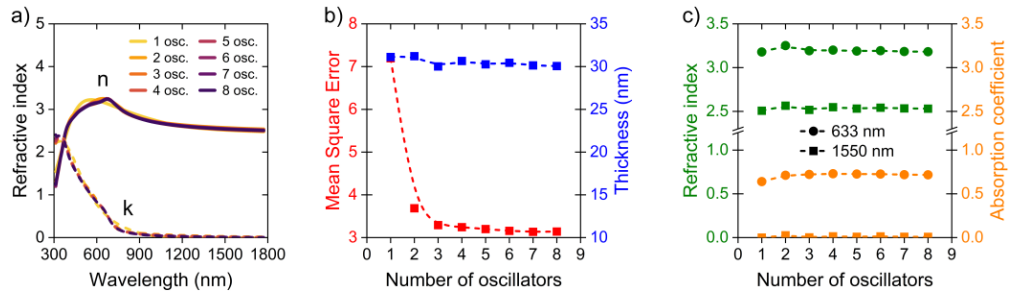

Fig. S15. Convergence test of a number of Tauc-Lorentz oscillators in the ellipsometry fitting of the crystalline $Sb_2S_3$ film deposited at 10 Hz, 0.5 J/cm$^2$, 100 mTorr and 1000 pulses (**29 nm**) and annealed at 300°C by 27°C/min ramp rate. a) Real (n) and imaginary (k) parts of the measured refractive index as functions of the wavelength for various numbers of Tauc-Lorentz oscillators used for the ellipsometry fitting. b) Mean square error of the ellipsometry fitting and the fitted thickness as functions of the number of oscillators used. c) The refractive index and the absorption coefficient at 633 nm and 1550 nm wavelengths as functions of the number of oscillators used. The refractive index, absorption, and thickness do not significantly change after 3 oscillators. b-c) Data are fitted by a b-spline.

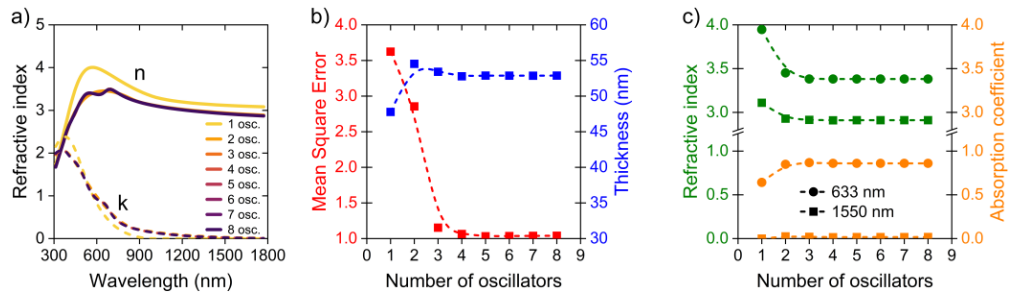

Fig. S16. Convergence test of a number of Tauc-Lorentz oscillators in the ellipsometry fitting of the crystalline $Sb_2S_3$ film deposited at 10 Hz, 0.5 J/cm$^2$, 100 mTorr and 2000 pulses (**60 nm**) and annealed at 300°C by 27°C/min ramp rate. a) Real (n) and imaginary (k) parts of the measured refractive index as functions of the wavelength for various numbers of Tauc-Lorentz oscillators used for the ellipsometry fitting. b) Mean square error of the ellipsometry fitting and the fitted thickness as functions of the number of oscillators used. c) The refractive index and the absorption coefficient at 633 nm and 1550 nm wavelengths as functions of the number of oscillators used. The refractive index, absorption, and thickness do not significantly change after 4 oscillators. b-c) Data are fitted by a b-spline.

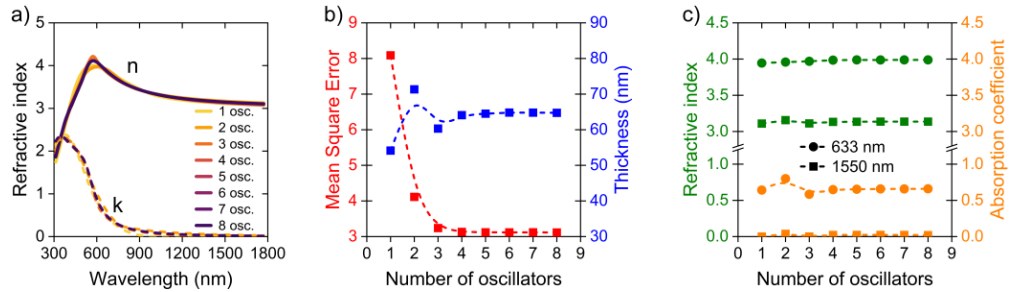

Fig. S17. Convergence test of a number of Tauc-Lorentz oscillators in the ellipsometry fitting of the crystalline $Sb_2S_3$ film deposited at 10 Hz, 0.5 J/cm$^2$, 100 mTorr and 3000 pulses (**89 nm**) and annealed at 300°C by 27°C/min ramp rate. a) Real (n) and imaginary (k) parts of the measured refractive index as functions of the wavelength for various numbers of Tauc-Lorentz oscillators used for the ellipsometry fitting. b) Mean square error of the ellipsometry fitting and the fitted thickness as functions of the number of oscillators used. c) The refractive index and the absorption coefficient at 633 nm and 1550 nm wavelengths as functions of the number of oscillators used. The refractive index, absorption, and thickness do not significantly change after 4 oscillators. b-c) Data are fitted by a b-spline.

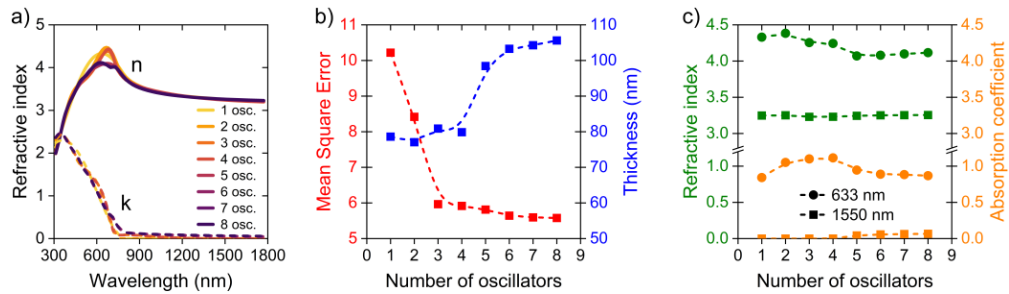

Fig. S18. Convergence test of a number of Tauc-Lorentz oscillators in the ellipsometry fitting of the crystalline $Sb_2S_3$ film deposited at 10 Hz, 0.5 J/cm$^2$, 100 mTorr and 4000 pulses (**116 nm**) and annealed at 300°C by 27°C/min ramp rate. a) Real (n) and imaginary (k) parts of the measured refractive index as functions of the wavelength for various numbers of Tauc-Lorentz oscillators used for the ellipsometry fitting. b) Mean square error of the ellipsometry fitting and the fitted thickness as functions of the number of oscillators used. c) The refractive index and the absorption coefficient at 633 nm and 1550 nm wavelengths as functions of the number of oscillators used. The refractive index, absorption, and thickness do not significantly change after 5 oscillators. b-c) Data are fitted by a b-spline.

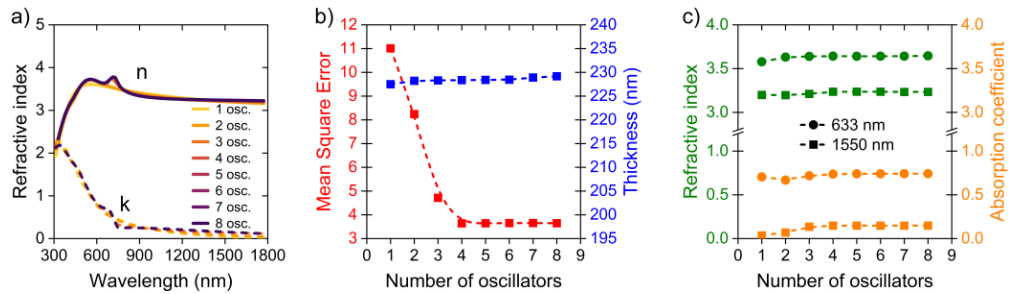

Fig. S19. Convergence test of a number of Tauc-Lorentz oscillators in the ellipsometry fitting of the crystalline $Sb_2S_3$ film deposited at 10 Hz, 0.5 J/cm$^2$, 100 mTorr and 10000 pulses (**264 nm**) and annealed at 300°C by 27°C/min ramp rate. a) Real (n) and imaginary (k) parts of the measured refractive index as functions of the wavelength for various numbers of Tauc-Lorentz oscillators used for the ellipsometry fitting. b) Mean square error of the ellipsometry fitting and the fitted thickness as functions of the number of oscillators used. c) The refractive index and the absorption coefficient at 633 nm and 1550 nm wavelengths as functions of the number of oscillators used. The refractive index, absorption, and thickness do not significantly change after 4 oscillators. b-c) Data are fitted by a b-spline.

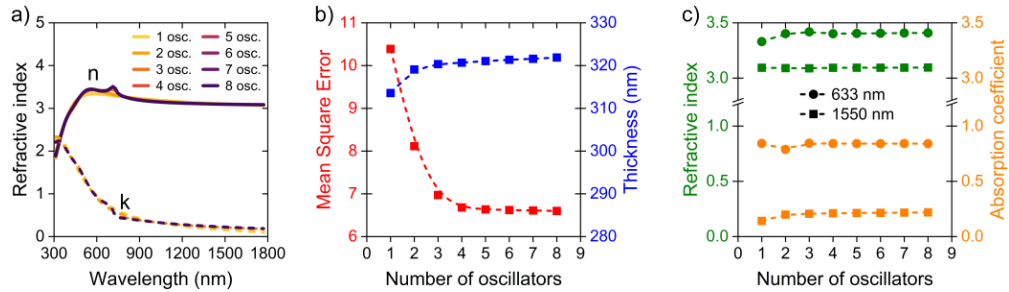

Fig. S20. Convergence test of a number of Tauc-Lorentz oscillators in the ellipsometry fitting of the crystalline $Sb_2S_3$ film deposited at 10 Hz, 0.5 J/cm$^2$, 100 mTorr and 15000 pulses (**386 nm**) and annealed at 300°C by 27°C/min ramp rate. a) Real (n) and imaginary (k) parts of the measured refractive index as functions of the wavelength for various numbers of Tauc-Lorentz oscillators used for the ellipsometry fitting. b) Mean square error of the ellipsometry fitting and the fitted thickness as functions of the number of oscillators used. c) The refractive index and the absorption coefficient at 633 nm and 1550 nm wavelengths as functions of the number of oscillators used. The refractive index, absorption, and thickness do not significantly change after 4 oscillators. b-c) Data are fitted by a b-spline.

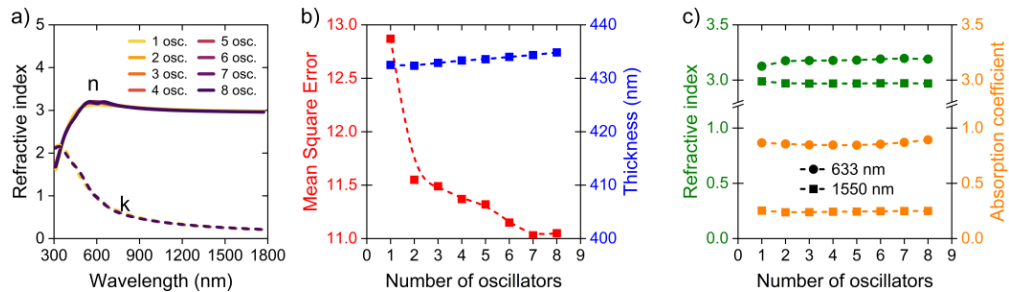

Fig. S21. Convergence test of a number of Tauc-Lorentz oscillators in the ellipsometry fitting of the crystalline $Sb_2S_3$ film deposited at 10 Hz, 0.5 J/cm$^2$, 100 mTorr and 20000 pulses (**493 nm**) and annealed at 300°C by 27°C/min ramp rate. a) Real (n) and imaginary (k) parts of the measured refractive index as the functions of wavelength for various numbers of Tauc-Lorentz oscillators used for the ellipsometry fitting. b) Mean square error of the ellipsometry fitting and the fitted thickness as functions of the number of oscillators used. c) The refractive index and the absorption coefficient at 633 nm and 1550 nm wavelengths as functions of the number of oscillators used. The refractive index, absorption, and thickness do not significantly change after 4 oscillators. b-c) Data are fitted by a b-spline.

| Thickness (nm) | Type | Amp – Amplitude | En – Position (eV) | C – Broadening | Eg – Bandgap (eV) |
|---|---|---|---|---|---|
| 16 | amorphous | 70.58 | 4.11 | 4.62 | 1.33 |
| | crystalline | 0.23 | 1.94 | 0.25 | 1E-4 |
| | | 92.85 | 2.55 | 1.67 | 1.63 |
| | | 342.92 | 2.93 | 0.89 | 3.24 |
| 29 | amorphous | 122.50 | 3.32 | 3.86 | 1.79 |
| | crystalline | 50.22 | 1.81 | 0.53 | 1.60 |
| | | 9.28 | 2.59 | 1.19 | 0.68 |
| | | 23.65 | 3.34 | 1.29 | 1.11 |
| 60 | amorphous | 161.12 | 2.82 | 2.83 | 1.93 |
| | crystalline | 7.62 | 1.85 | 0.33 | 1.35 |
| | | 3.16 | 2.46 | 0.63 | 1.07 |
| | | 165.90 | 2.81 | 1.41 | 2.89 |
| | | 28.64 | 2.92 | 1.85 | 0.63 |
| 89 | amorphous | 279.15 | 2.41 | 3.76 | 2.00 |
| | crystalline | 7.25 | 2.36 | 0.68 | 0.31 |
| | | 28.29 | 2.89 | 1.14 | 0.98 |
| | | 319.53 | 3.04 | 1.07 | 2.86 |
| | | 2.94 | 6.91 | 1E-6 | 0.37 |
| 116 | amorphous | 275.6 | 2.35 | 3.79 | 1.96 |
| | crystalline | 182.99 | 1.56 | 0.76 | 1.66 |
| | | 3.58 | 2.32 | 0.95 | 1E-4 |
| | | 27.44 | 2.61 | 1.59 | 1.45 |
| | | 154.82 | 2.73 | 2.12 | 2.07 |
| | | 2.03 | 3.41 | 0.60 | 1.39 |
| 141 | amorphous | 318.05 | 2.21 | 4.06 | 2.05 |
| | crystalline | 1.62 | 1.5 | 0.24 | 0.66 |
| | | 33.88 | 1.91 | 0.30 | 1.44 |
| | | 14.72 | 2.96 | 1.14 | 1E-4 |
| | | 462.16 | 2.99 | 0.97 | 2.88 |
| | | 52.16 | 2.38 | 0.52 | 1.88 |
| 264 | amorphous | 378.58 | 2.127 | 4.70 | 2.04 |
| | crystalline | 123.27 | 1.71 | 0.24 | 1.62 |
| | | 226.86 | 2.24 | 2.02 | 1.80 |
| | | 9.28 | 2.43 | 10.16 | 0.25 |
| | | 50.59 | 3.19 | 0.85 | 2.78 |
| 386 | amorphous | 376.47 | 1.97 | 4.38 | 2.09 |
| | crystalline | 256.09 | 1.69 | 0.10 | 1.91 |
| | | 53.22 | 2.40 | 1.25 | 1.76 |
| | | 1.79 | 3.46 | 0.95 | 1E-4 |
| | | 22.45 | 3.84 | 3.83 | 1E-4 |
| 493 | amorphous | 347.38 | 2.23 | 10.98 | 2.01 |
| | crystalline | 0.57 | 2.36 | 0.67 | 1E-4 |
| | | 0.67 | 2.58 | 0.70 | 1E-4 |
| | | 2.73 | 3.03 | 1.68 | 1E-4 |
| | | 23.20 | 3.63 | 2.98 | 1E-4 |

Tab. S3. Fitting parameters of ellipsometry measurement of amorphous and crystalline $Sb_2S_3$ films with thicknesses listed in the table.

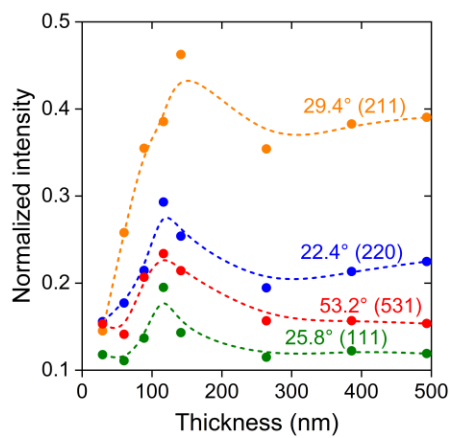

Fig. S22. Normalized intensities of the listed XRD peaks as functions of the film thickness. Data are fitted by a b-spline.

**References**

6. T. J. Whittles, T. D. Veal, C. N. Savory, A. W. Welch, F. W. S. Lucas, J. T. Gibbon, M. Birkett, R. J. Potter, D. O. Scanlon, A. Zakutayev, and V. R. Dhanak, "Core Levels, Band Alignments, and Valence-Band States in CuSbS2 for Solar Cell Applications," ACS Applied Materials & Interfaces **9**(48), 41916-41926 (2017).
7. NIST X-ray Photoelectron Spectroscopy Database, NIST Standard Reference Database Number 20, National Institute of Standards and Technology, Gaithersburg MD, 20899 (2000).
8. M. Schubert, T. Hofmann, C. M. Herzinger, and W. Dollase, "Generalized ellipsometry for orthorhombic, absorbing materials: dielectric functions, phonon modes and band-to-band transitions of Sb2S3," Thin Solid Films 455, 619–623 (2004).
32. W. Dong, H. Liu, J. K. Behera, L. Lu, R. J. H. Ng, K. V. Sreekanth, X. Zhou, J. K. W. Yang, and R. E. Simpson, "Wide Bandgap Phase Change Material Tuned Visible Photonics," Adv. Funct. Mater **29**, 1806181 (2019).
9. M. Delaney, I. Zeimpekis, D. Lawson, D. W. Hewak, and O. L. Muskens, "A New Family of Ultralow Loss Reversible Phase-Change Materials for Photonic Integrated Circuits: Sb2S3 and Sb2Se3," Adv. Funct. Mater. **30**, 2002447 (2020).
33. H. Liu, W. Dong, H. Wang, L. Lu, Q. Ruan, Y. S. Tan, R. E. Simpson, and J. K. W. Yang, "Rewritable color nanoprints in antimony trisulfide films," Sci. Adv. **6**, 7171–7187 (2020).
34. K. Gao, K. Du, S. Tian, H. Wang, L. Zhang, Y. Guo, B. Luo, W. Zhang, and T. Mei, "Intermediate Phase-Change States with Improved Cycling Durability of Sb2S3 by Femtosecond Multi-Pulse Laser Irradiation," Adv. Funct. Mater. **31**, 2103327 (2021).
35. Z. Fang, J. Zheng, A. Saxena, J. Whitehead, Y. Chen, and A. Majumdar, "Non-Volatile Reconfigurable Integrated Photonics Enabled by Broadband Low-Loss Phase Change Material," Adv. Optical Mater **9**, 2002049 (2021).
36. L. Lu, Z. Dong, F. Tijiptoharsono, R. J. H. Ng, H. Wang, S. D. Rezaei, Y. Wang, H. S. Leong, P. C. Lim, J. K. W. Yang, and Robert E. Simpson, "Reversible Tuning of Mie Resonances in the Visible Spectrum," ACS Nano **15**(12), 19722-19732 (2021).